\documentclass{aa}  
\usepackage{graphicx,color,subfigure}
\usepackage{txfonts}
\usepackage{verbatim}
\graphicspath{{./figures/}}
\usepackage{natbib,twoopt}
\usepackage[breaklinks=true]{hyperref} %% to avoid \citeads line fills

\hypersetup{
	colorlinks=true,   %% links colored instead of frames
	urlcolor=blue,     %% external hyperlinks
	linkcolor=blue     %% internal latex links (eg Fig)
}
\bibpunct{(}{)}{;}{a}{}{,} %% natbib format for A&A and ApJ

\makeatletter
\newcommand{\bibnote}[2]{\global\@namedef{#1note}{#2}}
\newcommand{\biblink}[2]{\global\@namedef{#1link}{#2}}
\makeatother
\makeatletter
\protected\def\stonyslink{%
	\def\hyper@linkstart##1##2{}\let\hyper@linkend\@empty}
\newcommandtwoopt{\citeads}[3][][]{%
	\href{http://adsabs.harvard.edu/abs/#3}%
	{\stonyslink \citealp[#1][#2]{#3}}%   %% Rutten, 2000
	\biblink{#3}{\href{http://adsabs.harvard.edu/abs/#3}{ADS}}}
\newcommandtwoopt{\citepads}[3][][]{%
	\href{http://adsabs.harvard.edu/abs/#3}%
	{\stonyslink \citep[#1][#2]{#3}}%     %% (Rutten 2000)
	\biblink{#3}{\href{http://adsabs.harvard.edu/abs/#3}{ADS}}}
\newcommandtwoopt{\citetads}[3][][]{%
	\href{http://adsabs.harvard.edu/abs/#3}%
	{\stonyslink \citet[#1][#2]{#3}}%     %% Rutten (2000)
	\biblink{#3}{\href{http://adsabs.harvard.edu/abs/#3}{ADS}}}
\newcommandtwoopt{\citeyearads}[3][][]{%
	\href{http://adsabs.harvard.edu/abs/#3}%
	{\stonyslink \citeyear[#1][#2]{#3}}%  %% 2000
	\biblink{#3}{\href{http://adsabs.harvard.edu/abs/#3}{ADS}}}
\makeatother
\begin{document}
\title{Asymmetric Capture into Neptunian 1:2 Resonance}
\author{Hailiang Li\inst{1,2}
	\and
	Li-Yong Zhou\inst{1}\fnmsep\inst{2}
	}
\authorrunning{Li \& Zhou}
\offprints{L.-Y. Zhou, \email zhouly@nju.edu.cn}
\institute{School of Astronomy and Space Science, Nanjing University, 163 Xianlin Avenue, Nanjing 210046, China
	\and
	Key Laboratory of Modern Astronomy and Astrophysics in Ministry of Education, Nanjing University, China
	}
\date{}

\abstract{The asymmetric resonance configuration characterized by the critical angle librating around centres other than $0^\circ$ or $180^\circ$, is found in the 1:N mean motion resonance. The asymmetric 1:2 resonance with Neptune is of particular interest because the two asymmetric islands seem to host different populations, and this might be a direct clue to understanding the early evolution of the Solar system. The asymmetry has been investigated from both observational and theoretical perspectives, but conclusions among studies vary widely. In this paper using toy models, we carefully designed a series of tests to systematically study the capture of planetesimals into the leading and trailing resonance islands. Although these tests may not reproduce exactly the real processes the Solar system experienced, they reveal some typical dynamics in the resonance capture. Since the real Twotinos have small to moderate inclinations, as the first attempt, we adopted in this paper planar models to investigate the mechanisms that may lead to asymmetric capture by the leading and trailing islands, including their size variation during the outward migration of Neptune, the stickiness of the leading island, and the migration slowdown effect. Particularly, we find that the ratio between the populations of the leading and trailing islands can be easily tuned by introducing the slowdown effect in the migration model, thus may be not a good tracer of the migration history. However, the eccentricity of objects trapped in two asymmetric islands may conserve some valuable information of the early evolution of the Solar system. }

\maketitle{}

\section{Introduction}
The distribution of Trans-Neptunian Objects (TNOs) in space bears major clues to the early evolution of the Solar system. Among TNOs, the resonant population is believed to have been captured into  mean motion resonances (MMRs) with Neptune \citepads[see e.g.][]{Malhotra1995, 2002MNRAS.336..520Z} during the late stage of the the Solar system formation when Neptune (and other planets) was migrating due to its scattering of a large number of planetesimals \citepads{1984Icar...58..109F}. Thus the existence of resonant populations, their physical and chemical characteristics, and their orbital distributions, as the consequences of resonance capture processes, are all of great interest and deserve thorough analyses. 

For a 1:N MMR with Neptune, except for the symmetric resonance configuration characterized by the resonance angle librating around $0^\circ$ or $180^\circ$, asymmetric resonance configurations in which the libration is around values other than $0^\circ$ or $180^\circ$ exist \citepads[e.g.][]{Frangakis1973, Malhotra1996}. The asymmetric configuration with the resonance angle librating between $0^\circ$ and $180^\circ$ is called ``leading resonance island'', while the other one who librates between $180^\circ$ and $360^\circ$ is ``trailing resonance island''. Theoretically, these two asymmetric islands are absolutely the
same in the point of view of dynamics under current planetary
configuration. As a matter of fact, the resonance could have complicated structure. For instance, a new resonance island may emerge at very high eccentricity in the planar circular restricted three-body model \citepads[e.g.][]{Beauge1994, Lan2019}, while in the non-planar case, very high inclination may introduce significant distortion to asymmetric islands \citepads[e.g.][]{Gallardo2006, Gallardo2020, Saillenfest2016, Efimov2020}. In the study of primordial planetesimal disk in the early stage of the Solar system, it is still appropriate to consider the two asymmetric islands as identical to each other, given that these planetesimals are anticipated to have relatively flat and circular orbits. The observations, on the contrary, show that the leading island is always superior in population. The most well-known example is the asymmetrical distribution of Trojan asteroids in the 1:1 MMR with Jupiter, where the Greek camp (around the Lagrange point L4) hosts about 90\% more asteroids than the Trojan camp (L5) according to the database of IAU Minor Planet Center (MPC)\footnote{https://minorplanetcenter.net/iau/lists/Trojans.html }. Similar asymmetry can be seen  apparently in the 1:N resonances with Neptune (see Table~\ref{tab:pop} in Section \ref{sec:asypop}). 

Many studies have devoted to the apparently asymmetric distribution of resonant TNOs in the 1:2 MMR with Neptune that are generally called ``Twotinos''. In the latest observation, \citetads{Chen2019} report 34 Twotinos, among which 17 are in the leading island, 8 in the trailing island, and the rest 9 in the symmetric configuration. Thus, the ratio of numbers of objects detected in the leading and trailing islands (hereafter L/T ratio) is $17/8\approx 2.1$. After taking into account the observation bias, \citetads{Chen2019} estimate a population of 4400 Twotinos with $H_r<8.66$ (diameter $\gtrsim$100\,km), among which 1600 are in the leading island and 1500 in the trailing island. They also deduce an overall L/T ratio within $0.25-1.86$ with a confidence of 95\%. We note that the observational bias favouring the detection of leading librator is due to the location of the trailing island which is in the direction of the Galaxy centre as seen from the Earth. Since Neptune has not moved a considerable distance after the TNOs were observed in quantity in last three decades, this observational bias may continue to blur the true distribution of TNOs in the asymmetric resonant islands in all the 1:N resonances with Neptune. 

\citetads{Chen2019} claim that the ``real'' Twotino populations in the two asymmetric islands are not so different from each other after accounting for the observational bias, which means the value of L/T ratio is almost 1. However, numerical simulations of the capturing of planetesimals into the resonance yield widely different L/T ratios. \citetads{Chiang2002} modeled the capturing of planetesimals into MMRs during the migration of Neptune and found that the L/T ratio of Twotinos was 0.91 and 0.30 respectively when the migration timescale is 10\,Myr and 1\,Myr. Later, \citetads{Murray2005} suggested that the migration should not be too fast, to avoid an overwhelming capture fraction into the trailing island. \citetads{Li2014} investigate the resonant capture of objects at different inclinations and obtain an L/T ratio $\sim$1.1 in low-inclination, where the migration timescale is 20\,Myr. \citetads{Pike2017} however find a leading-island-dominant result with L/T ratio of 2.24 in their numerical simulations of the evolution of outer Solar system following a specific version of Nice model \citepads{Brasser2013} where Neptune undergoes a high-eccentricity (0.3) phase. And the authors suggest that the orbital circularization of Neptune is not the reason of leading island enhancement. Adopting four different Neptunian migration models \citepads[as proposed by][]{Kaib2016} characterized by ``grainy slow'', ``grainy fast'', ``smooth slow'', and ``smooth fast'', \citetads{Lawler2019} calculate the capture efficiency of Twotinos, from which \citetads{Chen2019} derive the fraction of asymmetric Twotinos in the leading island in these migration models as 0.36, 0.54, 0.52 and 0.56. Translated to L/T ratio, these numbers are 0.56, 1.17, 1.08 and 1.27, respectively. 

So far, both in observational data and in numerical simulations, the asymmetric distribution of TNOs in the two resonant islands of 1:N MMRs is often recognised but the L/T ratio is reported in a wide range, and the possible mechanisms that may result in this asymmetry are still vague. 
We note that the non-gravitational effects, in particular the Yarkovsky effect, may contribute to the asymmetric distribution of asteroids in the 1:1 MMR with inner planets like the Earth and Venus \citepads[see e.g.][]{Zhou2019A&A, Xu2022A&A}, but they could hardly have any influence on TNOs due to the large distance to the Sun. Therefore, we have to search for the mechanisms leading to the asymmetric distribution of Twotinos mainly in the process of their being captured.  

\citetads{Murray2005} proposed three mechanisms that may contribute to the asymmetric capture of planetesimals into the 1:2 resonance. First, the size of the leading asymmetric resonant island shrinks while the trailing one expands as Neptune migrates outward. Consequently, an object  will have a larger probability to be trapped and stay around the trailing island than the leading one. Particularly, the leading island might even completely disappear if the migration is fast enough, so that the capture is totally dominated by the trailing island. The second mechanism is that the difference between the leading and trailing islands become less significant as the planetesimal's initial eccentricity increases, thus the trailing island's advantage in population decreases. And thirdly, an object's libration around the leading island is slower, thus an object spends longer time in the vicinity of the leading island than the trailing one before this object is finally trapped by either island. 

We devote this paper to studying the details of resonant capturing process of TNOs into the 1:2 MMR with Neptune through carefully designed numerical simulations and quantitative analyses, aiming at understanding the influences of migration model and initial conditions of planetesimals on the distribution of Twotinos in the leading and trailing islands. It is worth noting that the models adopted in this paper are often intentionally simplified to emphasize the dynamical effects of certain capturing process and we do not imply that Neptune migrated exactly in the way the models follow.  

The rest of this paper is arranged as follows. In Section 2, as the background of our investigations we introduce the dynamical structure of the 1:2 MMR. The numerical models adopted in this paper are also introduced in this section. Then the numerical simulations of the Neptunian migration and resonance captures are presented in Section 3, and the mechanisms that can cause the asymmetry between two asymmetric resonance islands are analysed. The discussion and conclusion are given in Section 4.

\section{Asymmetric resonance islands}

For a 1:N MMR in a planar restricted three-body model, the asymmetric resonance islands appear only when the eccentricity is larger than a specific value. For the 1:2 MMR with Neptune, the critical eccentricity is $\sim$0.04 and they disappear again after the eccentricity exceeds $\sim$0.95 \citepads[see e.g.][]{Lan2019}. 

\subsection{Dynamical structure of 1:2 resonance}
To give a sketch description of the structure of the 1:2 MMR with Neptune, we conduct some numerical simulations of the motion of test particles in the 1:2 MMR in the circular planar restricted three-body model where the primary bodies are the Sun and Neptune. 
The initial semimajor axis of test particles is set at the nominal location of the 1:2 MMR $a_0=2^{2/3}a_{\text N} \approx 47.622$\,AU, where the subscript `N' stands for Neptune. The initial mean anomaly is chosen as $M_0=0^\circ$. And the initial eccentricity $e_0$ and longitude of pericentre $\varpi_0$ are evenly distributed in the range $[0,0.6)$ and $[0^\circ,360^\circ]$. 
 
\begin{figure}[!htbp]
	\centering
	\resizebox{\hsize}{!}{\includegraphics{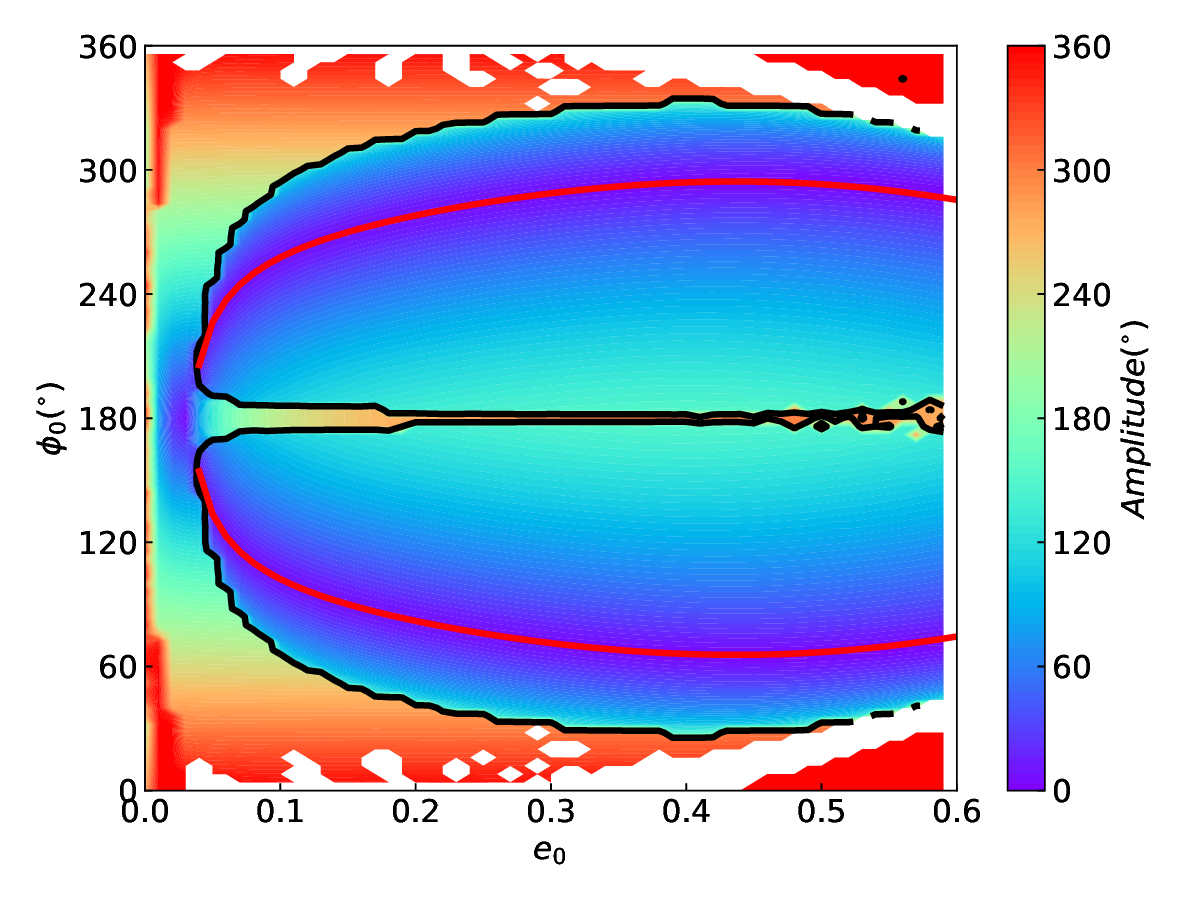}}
	\caption{The libration amplitude of resonance angle in numerical simulations of the circular planar restricted three-body model. The abscissa and ordinate are the initial  eccentricity $e_0$ and resonance angle $\phi_0$ respectively, and the libration amplitude is indicated by colour. The asymmetric islands are enclosed by black lines, and the red lines indicate the location of asymmetric resonance centre (see text).}
	\label{fig:AsySol}
\end{figure}

The equations of motion are then numerically integrated for 1\,Myr and we monitor the resonant angle $\phi=2\lambda-\lambda_N-\varpi$ of these test particles in the evolution, where $\lambda$ and $\lambda_N$ are the mean longitudes of the test particle and Neptune, respectively. In Fig.~\ref{fig:AsySol}, we summarize the libration amplitudes of $\phi$, which are indicated by colour. The blank space represents the non-resonance objects, and the black curves give roughly the boundary between symmetric and asymmetric resonances. 

The libration centre corresponds to the equilibrium solution to the equations of motion, which in turn can be obtained by determining the minimum of the perturbation function, and the red lines in Fig.~\ref{fig:AsySol} represent these equilibrium points. We note that the red lines fit very well the location of minimal libration amplitude of $\phi$ indicated by the bluest colour, which is obtained from numerical simulations of the motion. 

We set $M_0=0^\circ$ for test particles, and without loss of generality $\lambda_{\text N0}=0^\circ$ for Neptune. We see in Fig.~\ref{fig:AsySol} that a pair of asymmetric resonant islands starts from the bifurcation of symmetric resonance ($\phi_c=180^\circ$) when $e_0\gtrsim 0.04$. The libration centre of asymmetric resonance deviates away from $\phi_c=180^\circ$ as $e_0$ increases and the deviation reaches the maximum when $e_0\approx 0.44$, where $\phi_c\approx 70^\circ$ (leading island) and $290^\circ$ (trailing island). We stop our calculation at $e_0=0.6$ because in the Solar system a Twotino will cross Neptune orbit if its eccentricity exceeds 0.37 and Uranus orbit if it's larger than 0.6. But in the restricted three-body model, the two asymmetric resonant islands maintain until $e \approx 0.95$, after which they meet each other and merge again at $\phi_c=180^\circ$ \citepads{Lan2019}.

\subsection{Asymmetric population in 1:2 resonance} \label{sec:asypop}
Fig.~\ref{fig:AsySol} implies a symmetry between the leading and trailing asymmetric resonant islands, i.e. the two resonant configurations are dynamically identical to each other, and a same amount of objects may be expected in the two islands, if these objects do exist. However, contrary to this expectation, many 1:N (not only the 1:2) MMRs in the Solar system are found to have an apparent asymmetry in hosting objects between the leading and trailing island. For example, for the 1:1 resonance with Jupiter, \citetads{2007MNRAS.377.1393S} obtained an L/T ratio of $1.6\pm0.1$, while \citetads{Grav2011} estimated a value of $\sim$$1.4\pm0.2$. We list in Table~\ref{tab:pop} the number of objects located in some major 1:N resonances in the Solar system. In Table~\ref{tab:pop}, objects in the 1:1 resonance with Jupiter or Neptune are also known as Trojans, and their data is from the ``Lists of Trojan minor planets'' of IAU MPC on date July 2023\footnote{https://minorplanetcenter.net/iau/lists/Trojans.html }. For other 1:N resonances, the numbers in the table are obtained from our numerical simulations as follows. We downloaded from the Asteroids-Dynamic Site (AstDyS)\footnote{https://newton.spacedys.com/astdys} the orbital elements of objects locating nearby the 1:N resonance, and for each object we then generated 20 clone orbits within the observational error bars. These orbits were then integrated for 1\,Myr using {\it Swifter\_symba} \citepads{Levison2000} in the gravitational model consisting of the Sun and four giant planets. The resonant angles of clone orbits were monitored during the simulations and the proportion of them staying in the leading or trailing islands were calculated. Thus the decimal numbers in Table~\ref{tab:pop} imply that either some objects are located in the boundary zone of the resonance islands or the observation errors are large.

\begin{table}[!htbp]
	\centering
	\caption{The number of asteroids in the leading and trailing resonance islands and their ratio (L/T) in some major 1:N resonances. The data of 1:1 resonance with Jupiter or Neptune is from the IAU MPC, while other values are calculated based on our simulations, with the initial orbital elements obtained from the Asteroids-Dynamic Site. }
	\begin{tabular}{|c|l|l|l|c|}
		\hline
		Resonance & Leading  & Trailing  & Symmetric & L/T    \\ \hline
		Jup. 1:1  & 8275     & 4306      & N.A.      & 1.92         \\ \hline
		Nep. 1:1 & 27          & 4                   & \ \ 1                & 6.75         \\ 
		Nep. 1:2 & 32.4        & 9.8                 & 17.0                 & 3.31         \\ 
		Nep. 1:3 & \ \ 4.6     & 0.6                 & \ \ 8.1              & 7.67         \\ 
		Nep. 1:4 & \ \ 3.8     & 0.4                 & \ \ 0.8              & 9.50         \\ \hline
	\end{tabular}
	\tablefoot{For those 1:N (except for 1:1) MMRs hosting only a small number of detected objects, clone orbits are generated and  their orbital status are used to obtain the statistical numbers in this table (see text).}
	\label{tab:pop}
\end{table}

It is clear that the leading island possesses a larger population than the trailing one for all these 1:N MMRs. For Neptune's resonances, the L/T ratio is always larger than three.  \citetads{Chen2019} attribute the L/T ratio of observed Twotinos to observational bias, for similar reasons, all other small objects in the Neptune 1:N resonance are strongly affected by observational bias. Although current observations are not adequate enough to confirm whether the leading and trailing islands of the 1:2 resonance are identical or not in population, there are mechanisms in the dynamics that could result in different capture preferences in these islands \citepads[see, e.g.,][]{Murray2005}. Therefore, further theoretical investigation is necessary to explore the possibility of asymmetric capture of Twotinos and explain the discrepancies between the results obtained from different models \citepads[e.g.][]{Chiang2002,Li2014,Pike2017,Lawler2019}. This will provide an insight into the accuracy of our predictions regarding the actual distribution of Twotinos.

\subsection{Numerical model}
Since the leading and trailing islands are dynamically symmetric to each other under current configuration of the Solar system, the asymmetric populations must be the result of the capturing of objects into the resonances and/or the subsequent evolution in the resonances before planets attained their current orbits. We devote this paper to investigating the formation of Twotinos' population in the asymmetric islands. For this sake, we carry out numerical simulations of different scenarios of planetary migration and capturing of fictitious objects that represent the planetesimals in the disk.

The capturing of planetesimals into an MMR is a short-term process compared to the secular perturbations from planets other than Neptune, thus the effects of secular perturbations can be neglected if we focus on the dynamics of capturing. Moreover, the outcome of resonant capture sensitively depends on the initial eccentricity of planetesimals and migration rate, but not strongly on the inclination. In the simulation of \citetads{Li2014}, the L/T ratio does not vary much in the low-inclination ($i<20^\circ$) regime, while for high-inclination ($i>20^\circ$) the capture efficiency is very low thus not statistically significant. In addition, this paper focuses on the primordial capture process, when the inclination of planetesimals in the disk have not yet been excited. Therefore, in this paper, in numerical simulations of capturing of planetesimals into the 1:2 MMR with Neptune, a planar circular restricted three-body model was adopted, which consists of the Sun, Neptune on a circular orbit, and a number of zero-mass planetesimals. 

In our simulations, an artificial force was exerted on Neptune to make it migrate at a given rate $\dot{a}_\text{N}$. If $\dot{a}_\text{N}$ is constant, we have a linear migration. We note that an exponential migration is often applied in literature \citepads[e.g.][]{Malhotra1995}, in which the migration from an initial orbit $a_\text{i}$ to final one $a_\text{f}$ is characterized by a timescale $\tau$: $a(t)=a_\text{f}-(a_\text{f}-a_\text{i})\exp(-t/\tau)$. For cases in which $\dot{a}_\text{N}$ is not constant, the average rate $\langle\dot{a}\rangle$ is often used to characterize the migration and compare with the linear migration speed. It should be noted that the averaging should be calculated over the semimajor axis (but not over time as usual), because the planetesimals are usually distributed according to their distances from the Sun. %If the density gradient of the planetesimals' distribution is taken into account, there will be more capture at the inner position and the average rate may be somewhat higher. And 
During the migration, any migration rate change, such as that in grainy migration \citepads{Nesvorny2016} and the major ``jump'' of Neptune \citepads{Nesvorny2012}, increases the value of $\langle\dot{a}\rangle$ because high-rate stages weight more.

\section{Neptunian migration and resonance capture}

\subsection{An example case} \label{sec:basicmodel}
First of all, we run some numerical simulations to get an overall view of the planet's orbital migration and resonant capture. In these simulations, Neptune migrates from 26\,AU to 30\,AU at constant migration rate, and several migration rates from $\dot{a}_\text{N}=0.1$\,AU/Myr to 16\,AU/Myr were selected. We note that a wide range of migration rates have been adopted in previous works. For instance, for a Neptune migration of 7\,AU (from 23 to 30\,AU), the e-folding timesacles $\tau=0.1$\,Myr, 1\,Myr and 10\,Myr were tested in \citetads{Chiang2002}, and the longest $\tau$ was suggested. \citetads{Nesvorny2015a} tries several timescales from 1\,Myr to 100\,Myr and suggests $\tau \gtrsim 10$\,Myr is appropriate. The linear migration rates we adopted here are approximately equivalent to e-folding timescales of 0.2\,Myr ($\dot{a}_\text{N}=16$\,AU/Myr) to 35\,Myr ($\dot{a}_\text{N}=0.1$\,AU/Myr). 

\begin{figure}[!htp]
	\centering
	\resizebox{\hsize}{!}{\includegraphics{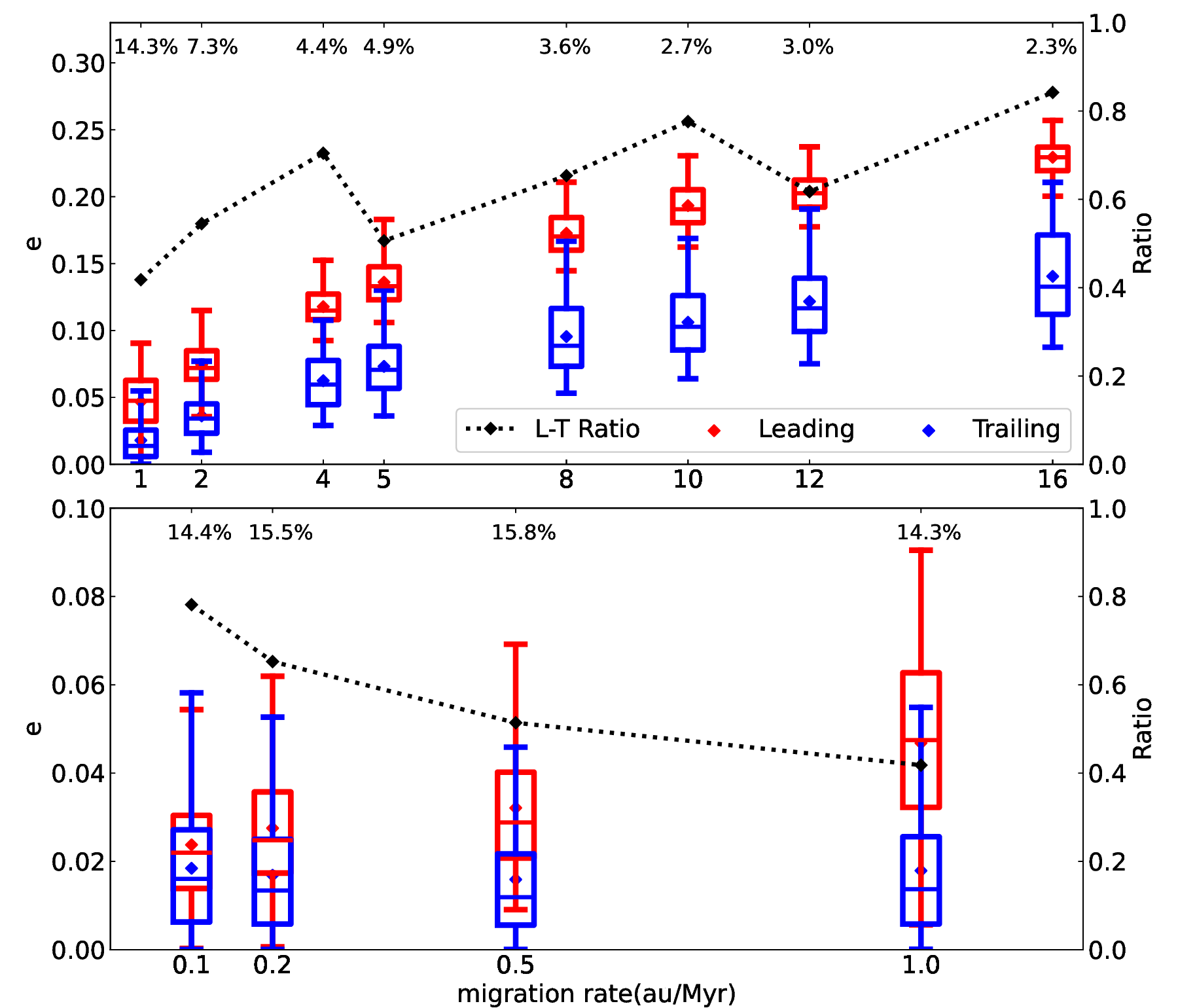}}
	\caption{The initial eccentricities of test particles captured by the leading island (red) and trailing island (blue) at different migration rates of Neptune. The box plots tell the minimum, maximum, median, and upper and lower quartiles as usual. The solid diamonds are the mean value of eccentricities. The L/T ratio is also plotted as solid circles connected by dotted curve (right ordinate). The total capture efficiencies (in per cent) are labelled on the top. The upper and lower panels are for fast and slow migration rate ranges respectively.}
	\label{fig:CapEff}
\end{figure}

For a certain $\dot{a}_\text{N}$, 10,000 test particles representing the planetesimals originally occupy evenly the region from 44.45\,AU to 46.03\,AU, corresponding to the nominal positions of the 1:2 MMR with Neptune when its semimajor axis is 28 and 29\,AU, respectively. After Neptune stops migration at 30\,AU, the simulation continues for another 0.2\,Myr and we check the resonant state of each test particle. The initial eccentricities of test particles are randomly distributed in $[0,0.3]$, the initial longitudes of pericentre and mean anomalies are randomly in $[0^\circ, 360^\circ]$. One may notice that the planetesimals are overexcited in eccentricity and might be quite different from the primordial planetesimal disk \citepads{1997AJ....114..841S, 2008ssbn.book..293K}. This is because we are not trying to reproduce the distribution of real planetesimals through these simulations, but to probe the dynamical mechanisms in theory. The results of simulations are summarized in Fig.~\ref{fig:CapEff}.

As Fig.~\ref{fig:CapEff} shows, the capture efficiency decreases from  $\sim$15\% (for $\dot{a}_\text{N}\leq1$\,AU/Myr) to $\sim$2\% (16\,AU/Myr) as the migration gets faster. %Considering that the eccentricity is randomly distributed in $[0,0.3]$ and Neptune only captures objects in a narrow range of eccentricity, capture efficiency is quite high when the migration is slow.
Although it is always smaller than 1, the L/T ratio shows a tendency of increasing to 1 when migration is either very fast or very slow. While much less test particles are captured by the leading island than the trailing one when migration rate is around 1\,AU/Myr. 

A slower migration is in favour of capturing test particles with lower eccentricities. In addition, the leading island tends to trap more highly eccentric particles than the trailing island, and the difference in eccentricity between them gets more significant when the migration is faster. Although the eccentricities of particles trapped by the trailing island are smaller than the leading island, the eccentricity range available for being captured is wider for the trailing island, especially for fast migration ($\dot{a}_\text{N}\gtrsim 4$\,AU/Myr). 

Except for the slow migration cases ($\dot{a}_\text{N}\lesssim 2$\,AU/Myr), the eccentricity ranges for objects being captured by two islands have only a little overlap, implying that not only we may expect more eccentric orbits in the leading island but also we can determine which island an object will join in by its eccentricity under a given migration rate. 

%Below, through more numerical simulations, we will check these mechanisms and investigate how the L/T ratio can be tuned.   

\subsection{Size variation of asymmetric islands}

\begin{figure}[!htp]
	\centering
	\resizebox{\hsize}{!}{\includegraphics{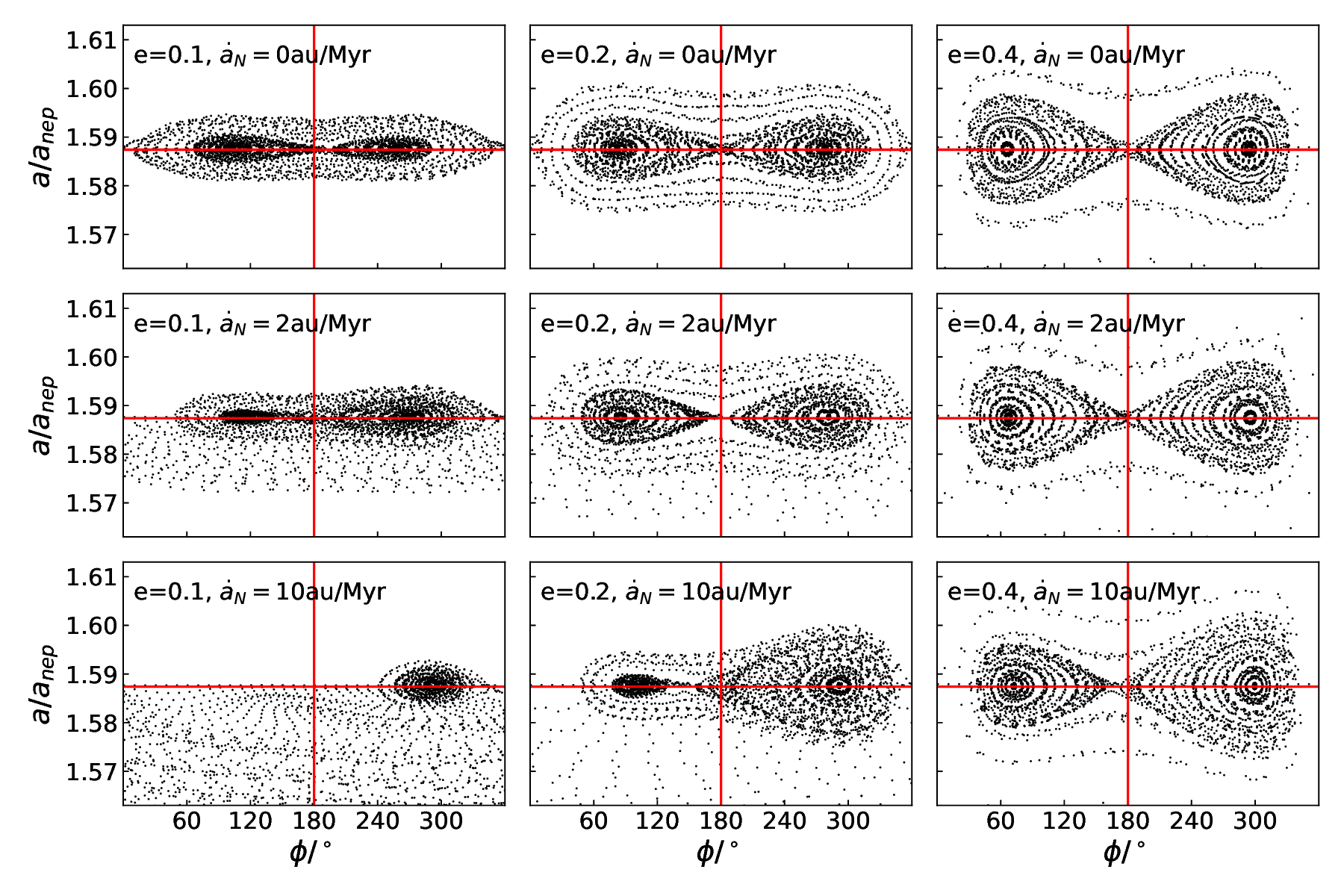}}
	\caption{Evolution of objects close to the 1:2 resonance, which demonstrates sizes of the resonance islands. The abscissa and ordinate represent the resonance angle $\phi$ and the semimajor axis (in Neptune's semimajor axis $a_{N}$), respectively. The planetesimal's eccentricity and Neptune's migration rate are indicated in each panel. For each particle, the time interval between two adjacent points is 1000 years. }
	\label{fig:2islands}
\end{figure}

We plot in Fig.~\ref{fig:2islands} some orbits of planetesimals with different initial eccentricities in the 1:2 MMR with Neptune migrating at different rates. In the simulations, Neptune was supposed to migrate outward from 29.0\,AU, and the initial semimajor axis of test particles was set at the nominal 1:2 MMR with Neptune, i.e. $a_0=46.03$\,AU. For these test particles, 36 different $\varpi_0$ are selected at $10^\circ$ intervals, while $M_0=0^\circ$ is fixed. The initial eccentricity of test particles and migration rates of Neptune is shown in each panel.

Apparently, the sizes of asymmetric islands vary, and the difference in size between two islands becomes more significant as  Neptune migrates faster but less significant as the initial eccentricity of planetesimals increases. 

The size of the resonance islands can be measured by the  extents of libration of $a$ and $\phi$ in the resonance. Thus a quantitative description of the size variation of resonance island can be obtained through a series of numerical simulations.

In these more detailed simulations, 180 different $\varpi_0$ from $0^\circ$ to $360^\circ$ at a $2^\circ$ interval and 20 different $e_0$ from 0.02 to 0.40 are set for test particles. Other initial conditions are similar to simulations presented in Fig.~\ref{fig:2islands}. Neptune was supposed to migrate outward from 29.0\,AU, and the initial semimajor axis of 3600 test particles was set at the nominal 1:2 MMR with Neptune, i.e. $a_0=46.03$\,AU. Several migration rates of Neptune from 0\,AU/Myr (no migration) to 15\,AU/Myr are chosen to test.

\begin{figure}[!htp]
	\centering
	\resizebox{\hsize}{!}{\includegraphics{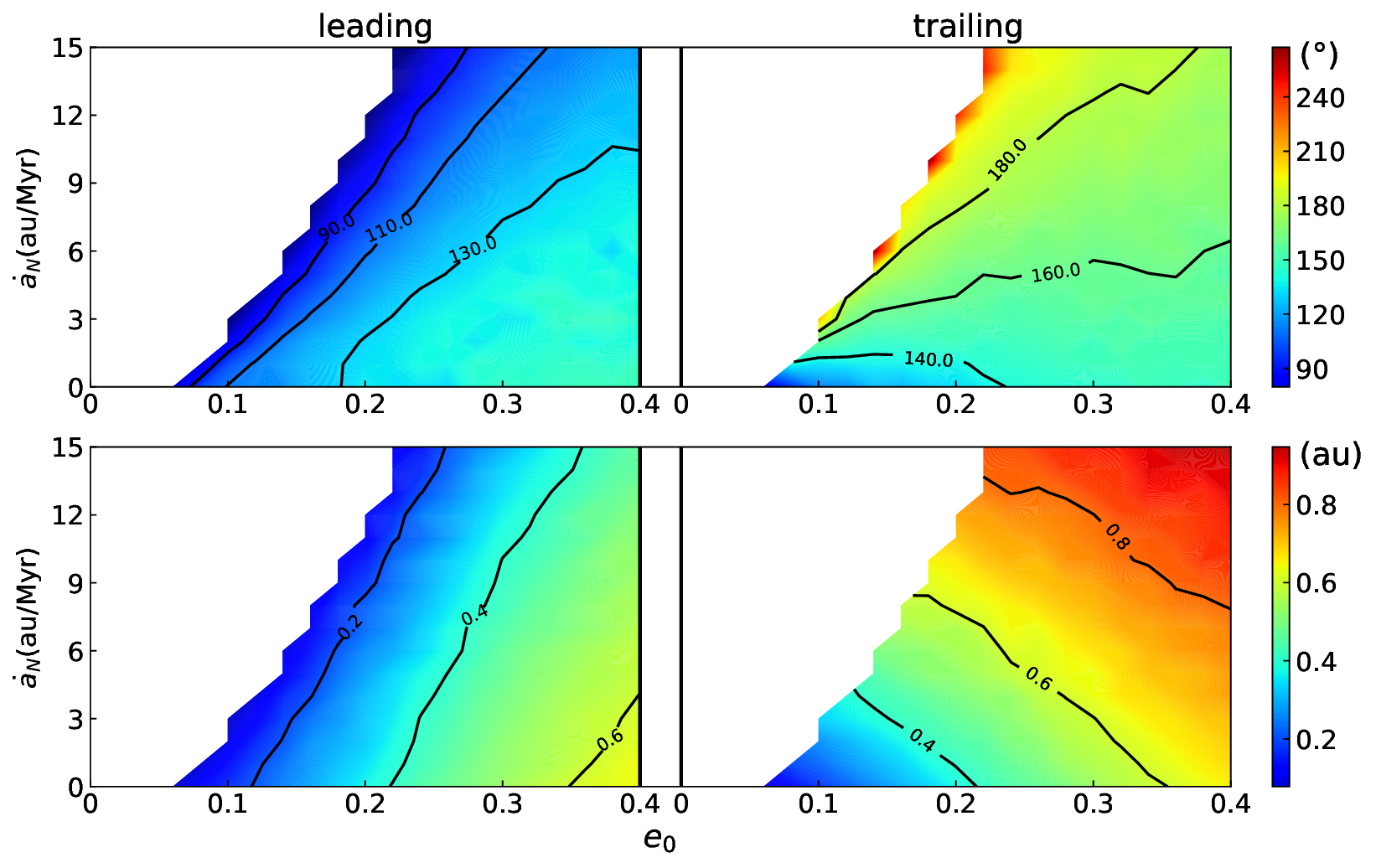}}
	\caption{The dependence of resonance island size on migration rate of Neptune ($\dot{a}_\text{N}$) and initial eccentricity of test particles ($e_0$). The maximal libration amplitude of resonance angle is shown by colour code in the upper panels, while the lower panels are for the libration of semimajor axis (see text). The left and right columns are for the leading and trailing island. }
	\label{fig:2islsize}
\end{figure}

From each run of simulation with given initial eccentricity and Neptune migration rate, we pick out the test particle that has the largest libration amplitude of resonance angle $\phi$. The libration amplitudes in $\phi$ and $a$ are then used as the indicator of the size of resonance islands, and we summarize the results in Fig.~\ref{fig:2islsize}.  

As shown in Fig.~\ref{fig:2islsize}, the asymmetric islands get bigger as the initial eccentricity increases. For $\dot{a}_\text{N}=0$, the island extends to have a $\phi$ amplitude of $\sim$140$^\circ$ and an $a$ amplitude of $\gtrsim$0.6\,AU.  The leading island shrinks as the migration rate increases. For example, for a given initial eccentricity $e_0=0.2$, the $\phi$ libration decreases from $\gtrsim$130$^\circ$ at $\dot{a}_N=0$ to $\lesssim$90$^\circ$ at $\dot{a}_N=9$\,AU/Myr, and the leading island even totally vanishes for larger migration rate.  On the contrary, the trailing island expands as Neptune migrates faster. Also for $e_0=0.2$, the libration amplitude of $\phi$ increases from $\lesssim$140$^\circ$ to $\gtrsim$180$^\circ$ as $\dot{a}_N$ increases from 0 to 9\,AU/Myr. The trailing island's expanding can also be found at the libration of semimajor axis $a$ in the lower panels in Fig.~\ref{fig:2islsize}. We note that the blank region at the corner of small eccentricity $e_0$ and fast migration $\dot{a}_N$ in Fig.~\ref{fig:2islsize} indicates that the resonant motion there is forbidden, i.e. in fast migration, only those planetesimals with eccentricity higher than a specific value can be captured, into either the leading or trailing island. 

Thus far we confirm here that the size variation of the asymmetric resonance islands is in favour of more planetesimals being trapped by the trailing island during the outward migration of Neptune. The two asymmetric islands host planetesimals with different eccentricities, and overall the difference in islands' size diminishes gradually as the eccentricity increases.

\subsection{Stickiness of leading island}

As \citetads{Murray2005} suggested, the only-one mechanism that might be in favour of leading island's capturing is that objects will be slowed down and thus aggregate around the leading island due to its shrinkage during the orbital migration. This effect is called {\it stickiness} here and we check this idea quantitatively as follows.  We have to note that the word ``stickiness'' here just describes the phenomenon that an orbit spends longer time around one resonance island than around the other one during the planetary migration, and it has nothing to do with the usual ``resonance stickiness'' that happens around resonances without planetary migration \citepads[see e.g.][]{Malyshkin1999, Lykawka2007, Sun2009, Bannister2016, Gallardo2018}. 

Again in our simulations, Neptune migrates outward from 29.0\,AU, and 10,000 test particles are supposed to locate in the nominal 1:2 resonance at 46.03\,AU. The initial eccentricities of them are in $[0, 0.3]$ and angular elements ($\varpi$ and $M$) are uniformly distributed in $[0^\circ, 360^\circ]$, respectively. Three migration rates of Neptune, 0.2\,AU/Myr, 1\,AU/Myr and 2\,AU/Myr, are tested. 

\begin{figure*}[!htp]
	\centering
	\includegraphics[width=\textwidth]{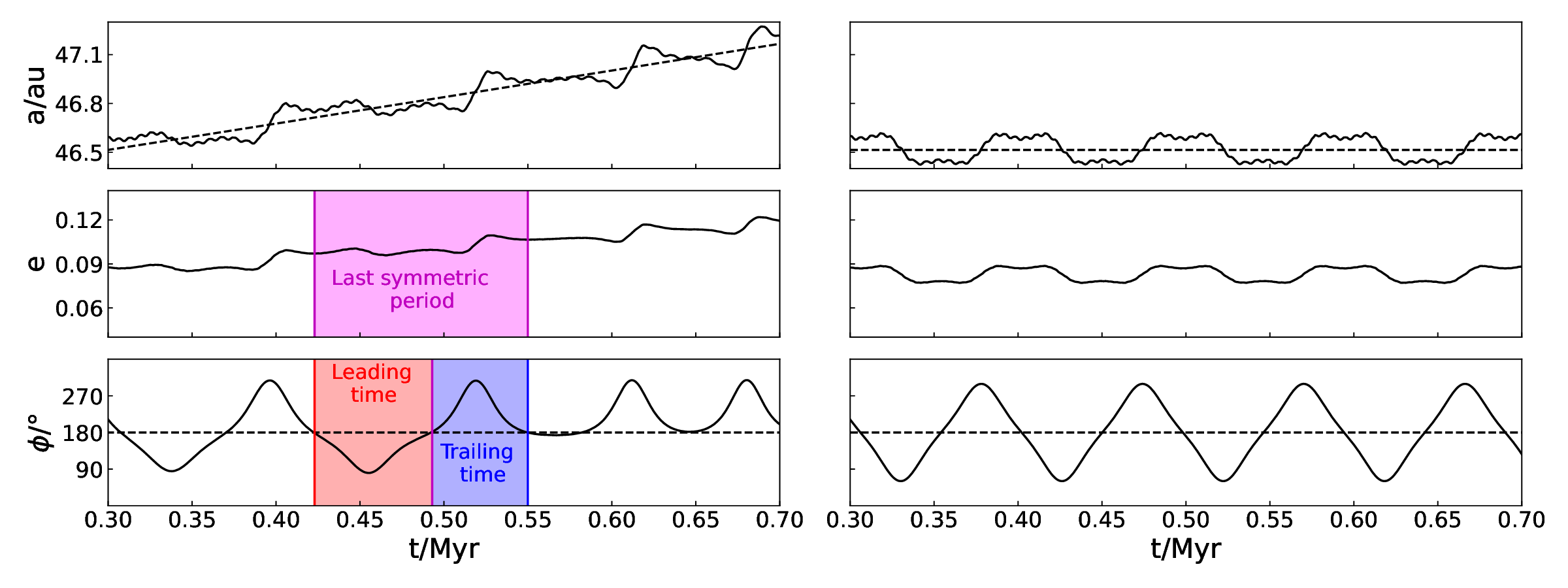}
	\caption{Evolution of a test particle around the transferring from symmetric to asymmetric resonance. The left column is for Neptune migrating at a rate 1\,AU/Myr while the right column for comparison is for the same test particle but the migration stops at 0.3\,Myr. From top to bottom, the semimajor axis $a$, eccentricity $e$ and resonant angle $\phi$ of the test particle are shown in solid lines. The dashed lines in the top panel represent the nominal semimajor axis of the 1:2 resonance. The magenta area indicates the duration of the last symmetric cycle, while the red and blue areas indicate the time intervals around the leading and trailing island, respectively. }
	\label{fig:timeratio}
\end{figure*}

A test particle of small eccentricity in the simulation may start from the symmetric resonance with resonant angle $\phi$ librating around $180^\circ$, and its libration amplitude grows accompanied by an increasing eccentricity as Neptune migrates outward. Then the test particle may evolve into an asymmetric resonance island, but before that it has experienced some complete symmetric resonance cycles. We show in Fig.~\ref{fig:timeratio} an example of such typical evolution. The stickiness  effect of the leading island can be recognized from the bottom left panel of Fig.~\ref{fig:timeratio}. Below $180^\circ$, the resonant angle $\phi$ oscillates at a relatively slower pace than above $180^\circ$, so that it stays a longer time around the leading island (red area) than around the trailing island (blue area), although the latter one is wider (which can be inferred from the larger amplitude of $\phi$).

\begin{figure}[!htp]
	\centering
	\resizebox{\hsize}{!}{\includegraphics{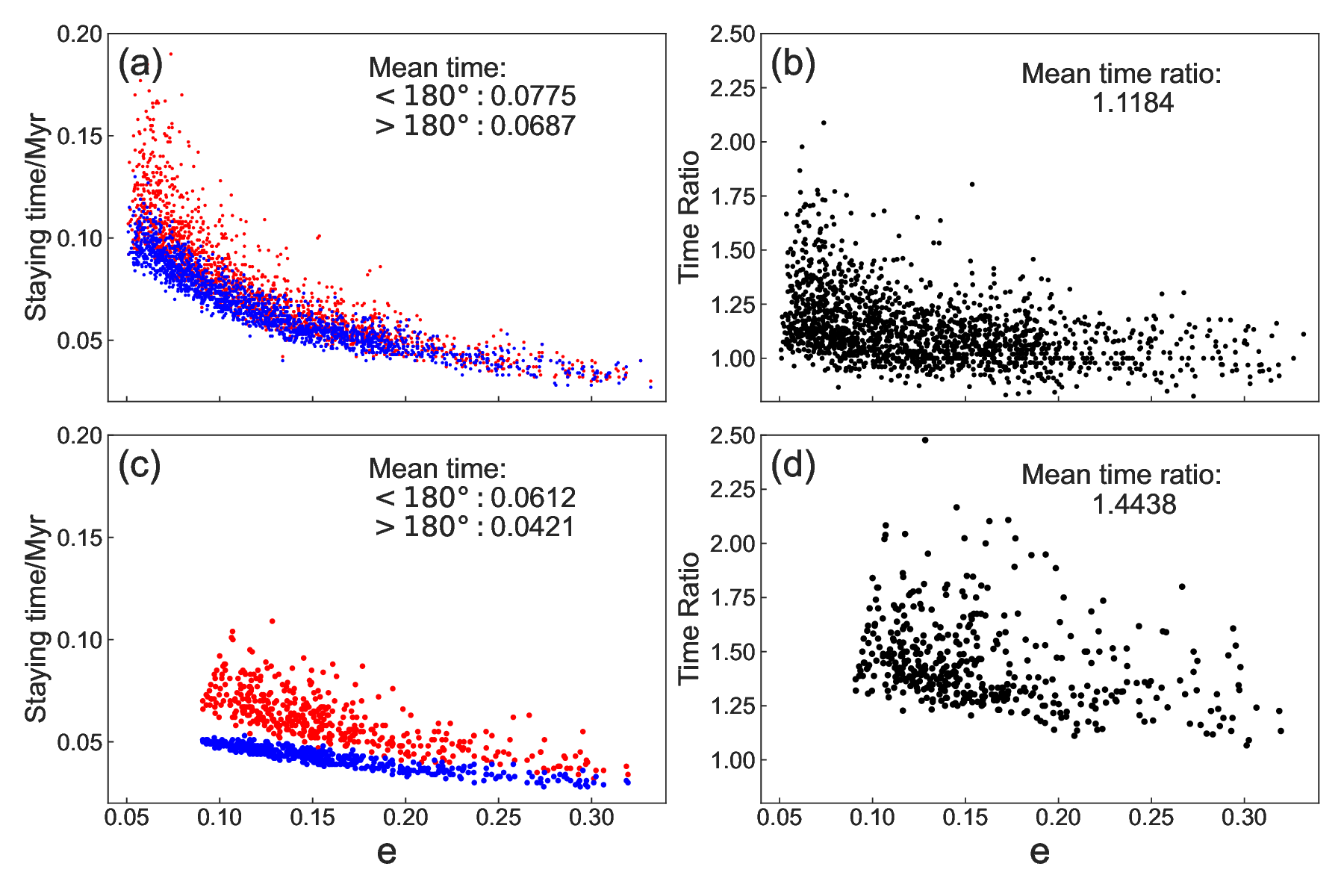}}
	\caption{The stickiness effect of the leading island. The times that an object staying on both sides of $\phi=180^\circ$ when it is transferring from symmetric to asymmetric resonances (see text and Fig.~\ref{fig:timeratio}) are plotted in the left two panels, and the ratio of the two time durations are shown in the right panels. The abscissa is the mean eccentricity in the last cycle before being captured by the asymmetric resonance. Red and blue dots represent the time spending in the leading and trailing side, respectively. The upper and lower panels represent the results from the simulations of migration rate $\dot {a}_\text{N}=0.2$\,AU/Myr and $\dot {a}_\text{N}=2$\,AU/Myr, respectively. In the former simulation, 1833 out of 10,000 initial objects are shown, and the number is 435 for the latter.}
	\label{fig:stickiness}
\end{figure}

In order to quantify the stickiness effect of leading island, for each test particle in our simulations that experiences the transferring from symmetric to asymmetric resonance, we examine the last symmetrical libration period before it's captured by either one of the asymmetric islands, and record the times that it stays on each side of $180^\circ$ (the red and blue shadowed areas in Fig.~\ref{fig:timeratio}), as well as its mean eccentricity during this period. We summarize the results of the cases with migration rates 0.2\,AU/Myr and 2\,AU/Myr in Fig.~\ref{fig:stickiness}.

Apparently in Fig.~\ref{fig:stickiness}, 80\% of test particles in the slow migration model (0.2\,AU/Myr) and 100\% test particles in the fast migration model (2\,AU/Myr) stay longer around the leading island ($\phi < 180^\circ$) in their last symmetric periods. Take the slower migration model as an example, test particles spend $\sim$12\% longer time around the leading side than in the other side, even though the leading island is smaller (see Fig.~\ref{fig:2islands}). Assuming the probability of being captured by either one of the resonance islands is proportional to the time an object spending around the island before the capture, we may expect 12\% more objects will be trapped in the leading island. Also can be found is that the ratio between the two times decreases as the eccentricity of test particles increases, implying that the stickiness effect of the leading island abates. In fact, when eccentricity is large, both islands expand (see Fig.~\ref{fig:2islands}), and the difference in island size caused by Neptune migration becomes less significant. A faster migration amplifies the asymmetry between two islands, just as it enhances their difference in size. 

So far, we find that the size variation of asymmetric islands favours the capture of planetesimals into the trailing island, while the stickiness effect favours the leading island, and these two opposite effects are both enhanced if the initial eccentricity of planetesimals is small or the migration is fast.

\subsection{Migration slowdown effect}
When Neptune migrates outward, the leading island shrinks while the trailing island expands, and the faster the migration, the greater the size difference between two islands. Certainly, if the migration decelerates, the difference between the asymmetric islands diminishes, until the migration stops and two islands regain the same size. During this migration slowdown process, the leading island may capture extra objects that were wandering in its close neighbouring region as it expands, while some objects may escape from the trailing island. In this paper, this phenomenon is referred to as {\it slowdown effect}, by which we mean the deceleration of Neptune's outward migration. We note that the migration may completely stops or even reverses (i.e. migrates inward to the Sun) if the deceleration is large enough. Considering the stochasticity of scattering events between Neptune and planetesimals that drives the migration, such temporary reverse migration may not be uncommon, but of course the most common scenario must be that the migration slows down due to the gradual depletion of planetesimals. This migration slowdown will change the L/T ratio, and we investigate this effect through some numerical simulations, as follows. 

\subsubsection{Numerical simulations}

Neptune was set to migrate from a semimajor axis of 29\,AU at an initial rate of $\dot {a}_\text{N1} = 5$\,AU/Myr. The migration was changed to a new speed $\dot {a}_\text{N2}$ after 0.2\,Myr when Neptune reaches 30\,AU. All test particles were placed at 47.62\,AU, exactly the nominal position of the 1:2 resonance with Neptune at 30\,AU. In such an arrangement, when the migration slowdown occurs, test particles are just ``touched'' by the resonance thus have not experienced the eccentricity evolution ``inside'' the resonance. The initial eccentricities were set from 0 to 0.3, with an interval of 0.015, thus 20 groups in total. For each group of test particles with a given eccentricity, the longitude of pericentre $\varpi$ is uniformly chosen from $0^\circ$ to $360^\circ$ with a $2^\circ$ interval, and the mean anomaly $M=0^\circ$ for all test particles. The total number of test particles are then 3600.

Selecting some migration rates $\dot{a}_\text{N2}$, we continued the simulations up to 0.4\,Myr, and during the simulation the libration of resonant angle $\phi$ for every object was monitored to check whether it is captured by either of the asymmetric resonant islands. The results are summarized in Table~\ref{tab:rst-brk}. Again, we must remind that the migration rate of Neptune in reality does not change to another value instantaneously. We use these ``unphysical'' tests to exaggerate and emphasize the effect of the migration rate changing.  

\begin{table}[htbp]
    \caption{The number of asymmetric resonance captures at different migration rate changings. Neptune's migration rate changes from $\dot {a}_\text{N1} =5$\,AU/Myr to $\dot {a}_\text{N2}$ listed in the first column (see text). The numbers of test particles captured by the leading and trailing islands, the overall capture efficiency, and the L/T ratio are in the second to the fifth column, respectively. }
    \centering
    \begin{tabular}{|c|c|c|c|c|c|}
    \hline
    Case & $\dot {a}_\text{N2}$ & Leading  & Trailing & Capture & L/T \\ 
      \# & (AU/Myr) & island & island & efficiency  & ratio \\ \hline
      1 & 6.00    	& \ \ 0  & \ 38 	& 1.06\% & 0.00\\ 
      2 & 5.00     	& \ \ 2  & \ 82 	& 2.33\% & 0.02 \\ 
      3 & 2.50   	& \ 27 	& 260 	& 7.97\% & 0.10 \\ 
      4 & 1.00     	& \ 99 	& 443 	& 15.1\% & 0.22 \\ %\hline
      5 & 0.50   	& 181 	& 408	& 16.4\% & 0.44 \\ %\hline
      6 & 0.25 		& 184 	& 135 	& 8.86\% & 1.36 \\ %\hline
      7 & 0.10   	& 206 	& \ 36 	& 6.72\% & 5.72 \\ %\hline
      8 & 0.00    	& 239	& \ 37 	& 7.67\% & 6.46 \\ %\hline
      9 & $-1.00$	& 328	& \ \ 0 	& 9.11\% & $\infty$ \\ \hline

    \end{tabular}

    \label{tab:rst-brk}
\end{table}

Migrating at a constant rate $\dot {a}_\text{N1} =\dot {a}_\text{N2} = 5$\,AU/Myr, Neptune captures only a small fraction ($2.33\%=84/3600$) of test particles into the 1:2 resonance, and most of them are in the trailing resonant island rather than the leading island (82 versus 2). If the migration accelerates to $\dot {a}_\text{N2} =6$\,AU/Myr, the capture efficiency drops down to 1.06\% and all the 38 particles are trapped in the trailing island (see Table~\ref{tab:rst-brk}). 

However, the capture efficiency, particularly for the leading island, increases if the migration decelerates ($\dot{a}_\text{N2} < \dot{a}_\text{N1}= 5$\,AU/Myr) as shown in Table~\ref{tab:rst-brk}. As a result, the L/T ratio increases significantly from 0.02 (for constant migration rate 5\,AU/Myr) to 6.46 (for a complete stop of migration, $\dot{a}_\text{N2}=0$). We also test a reverse migration by setting ${a}_\text{N2} = -1$\,AU/Myr, in which all test particles are found trapped in the leading island. 

Clearly, the change of Neptune's migration rate can influence substantially the outcome of resonant captures. In fact, due to Neptune's outward migration, the leading resonant island could be much smaller than the trailing island (Fig.~\ref{fig:2islands}), thus majority of test particles are captured by the trailing island. But, at the moment when the migration is ``interrupted'' by a slower migration rate (deceleration), the leading island expands while the trailing island shrinks relatively. Consequently, the leading island may devour those objects formerly wandering around or passing by the island. On the contrary, some objects formerly trapped in the trailing island may fall out from the island as it shrinks. Both of the above processes may contribute to the increase of the L/T ratio of captured test particles. 

Here we should revisit the resonance capturing process described in Section \ref{sec:basicmodel}, where the results of being captured into the two asymmetric islands (Fig.~\ref{fig:CapEff}) were calculated in 0.2\,Myr after Neptune's migration stopped, i.e. the slowdown effect discussed above have taken effects in fact in those simulations. If we regard the difference between two migration rates  applied in one simulation ($\Delta\dot{a}_\text{N} =\dot{a}_\text{N1}-\dot{a}_\text{N2}$) as a rough estimate of the strength of slowdown effect, we will find that the trends of L/T ratio changing with $\Delta\dot{a}_\text{N}$ in Fig.~\ref{fig:CapEff} and in Table~\ref{tab:rst-brk} agree with each other. However, the values of L/T ratios in these two numerical experiments, for example the Case 8 in Table~\ref{tab:rst-brk} and the case of $\dot{a}_\text{N}=5$\,Au/Myr in Fig.~\ref{fig:CapEff}, should not be compared directly, because different settings have been used in the simulations, particularly different initial populations of test particles. 

We note that the stickiness of the leading island contributes to the L/T ratio in this migration slowdown process, because the stickiness effect accumulates more particles near the leading island, providing a greater probability for them of being captured by the expanding island at the moment of migration deceleration.

\subsubsection{Resonance capturing process}

\begin{figure}[!htp]
\centering
 \resizebox{\hsize}{!}{\includegraphics{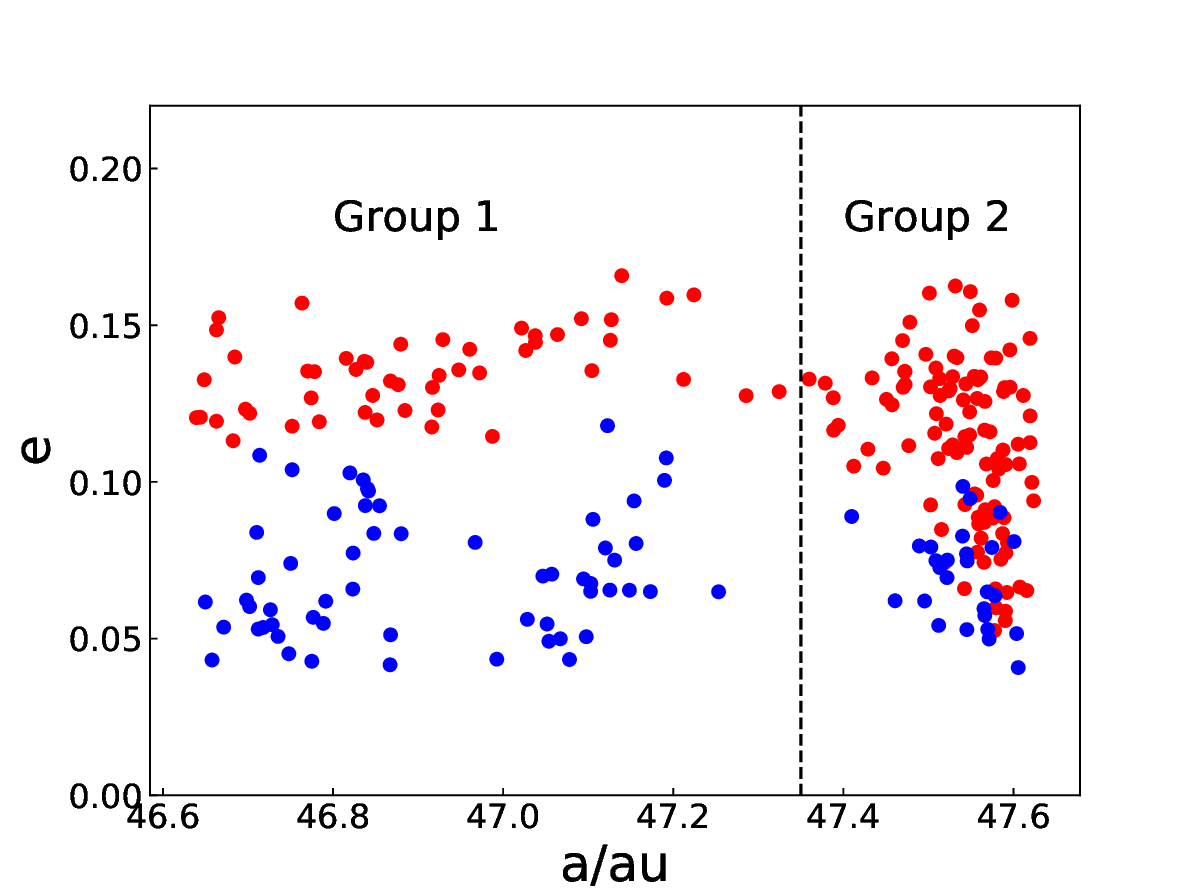}}
\caption{Distribution of test particles that are captured by the sweeping 1:2 resonance, The resonance sweeps through this region as Neptune migrates at a constant rate 5\,AU/Myr, and the migration is stopped when the nominal resonance reaches 47.62\,AU. The abscissa and ordinate are for the initial semimajor axis and eccentricity of test particles. Red and blue indicate that an object is captured by the leading and trailing resonance island, respectively. The dashed line at 47.35\,AU separates test particles into two groups (see text). }
\label{fig:brakeff}
\end{figure}

By setting all test particles to have the same initial semimajor axis (47.62\,AU) corresponding to the nominal 1:2 resonance exactly when Neptune changes its migration rate, we have shown how a prompt acceleration/deceleration of migration may influence the capture of objects to asymmetric resonant islands. However, the resonance capture must be rather a continuous process than a momentary event. To further clarify how the slowdown effect may influence the resonance capturing process, we repeat the simulation as Case 8 in Table~\ref{tab:rst-brk}, but this time 10,000 test particles are distributed evenly in a spatial range of 1\,AU before 47.62\,AU. Thus the 1:2 resonance will sweep across these test particles as Neptune migrates at a constant rate of 5\,AU/Myr, until the migration is removed when the nominal resonance reaches the outer edge of the test particles belt. After that, the evolution of test particles was followed for another 0.2\,Myr and we check their resonance configurations. Through these simulations, we may find out how the ``capture process'' by a sweeping resonance coordinates the destinations of test particles. We illustrate our results in Fig.~\ref{fig:brakeff}, where we plot the initial semimajor axes and eccentricities of test particles that are captured by one of the asymmetric islands. 

In Fig.~\ref{fig:brakeff}, the test particles form apparently two groups, separated from each other by semimajor axis $\sim$47.35\,AU, closely around which almost no successful capture into the resonance occurs. The Group 1 on the left side hosts 109 objects, less than half of which are captured into the leading resonance island, making an L/T ratio 0.88 ($=51/58$). On the right-hand side, 132 test particles make the Group 2 in which obviously more objects are captured by the leading island and the L/T ratio is 3.89 ($=105/27$). Since the initial semimajor axes of test particles are evenly distributed in the 1\,AU range from 46.62 to 47.62\,AU, the original population (located from 46.62 to 47.35\,AU) from which Group 1 in the left arise is about four times the initial population of Group 2 in the right, and the latter capture efficiency is much higher than the former. 

In our model, the 1:2 resonance sweeps through the region from 46.62 to 47.62\,AU at a constant rate and the migration stops when the resonance just passes the right edge of test particles belt. Thus  in Group 1 the latest-captured objects with initial semimajor axis $\sim$47.35\,AU have met the resonance 0.034\,Myr before the migration stops (it takes 0.034\,Myr for the nominal resonance to travel 0.27\,AU from 47.35\,AU to 47.62\,AU when Neptune migrates at 5\,AU/Myr). Meanwhile, we find  in our simulations that the typical resonant libration period around the asymmetric island is $\sim$0.07\,Myr, thus those objects in Group 1 have finished at least half a resonance libration period before the migration stops. Therefore, when the migration deceleration occurs (migration rate changes from 5\,AU/Myr to 0\,AU/Myr in this case), those objects in Group 1 have already been deeply in the resonance, and they are less affected by the deceleration. On the contrary, for objects in Group 2, the capture process is ongoing when the migration stops, and the deceleration affects significantly their final resonance configuration. Even for those objects on the right edge of the region (47.62\,AU), where the nominal resonance has just arrived as the migration is removed, the slowdown effect is obviously ``felt''.  As a result, the L/T ratio can be increased significantly by the slowdown effect of migration.  

Another feature that can be seen in Fig.~\ref{fig:brakeff} is that the leading island captures objects from high-eccentricity orbits more easily, which is consistent with the results in Fig.~\ref{fig:CapEff} and Fig.~\ref{fig:stickiness}. And, if the high L/T ratio was a result of such migration-slowdown mechanism, besides more eccentric orbits would be found in the leading island, we would also expect that the leading island capture objects later and from the region of more distance from the Sun when the deceleration of migration begins to take effect. If the chemical composition and physical property of planetesimals change radially with their distances to the Sun \citepads[see e.g.][]{Wong2016, Morbidelli2020}, we would argue that the objects now in the two asymmetric resonance islands may be different in colour, albedo, size, and composition, although such a discrepancy has not been observed yet \citepads{Chen2019}. 

The resonance capture is a process taking a certain period to complete, and the difference between two groups in Fig.~\ref{fig:brakeff} implies that its outcome can be altered if the migration changes during the period. In fact, the resonance begins its trapping of objects before the nominal position reaches the objects.

\begin{figure}[!htp]
	\centering
	\resizebox{\hsize}{!}{\includegraphics{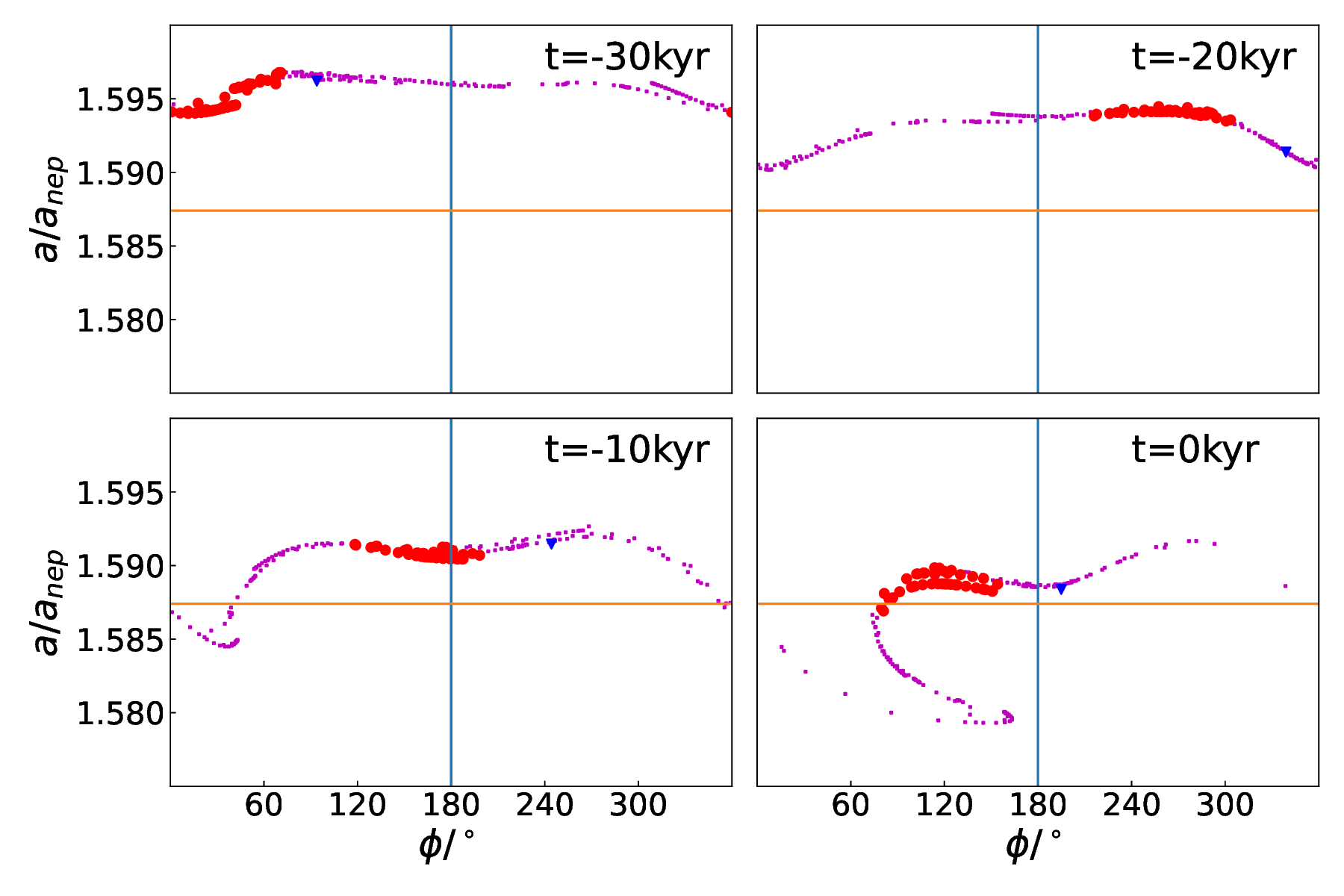}}
	\caption{Test particles' distribution on $(\phi, a)$ plane as the 1:2 resonance is approaching. Four snapshots at different moments, with $t=0$\,kyr representing exactly when the nominal resonance reaches the initial position of test particles, are given in four panels.  Objects later captured by the leading and trailing island are painted red and blue, while those purple ones are not captured by any asymmetric resonance. The horizontal line stands for the nominal position of 1:2 resonance.}
	\label{fig:precap}
\end{figure}

We arbitrarily choose again from Case 8 in Table~\ref{tab:rst-brk} the group of test particles with eccentricity $e_0=0.09$ and plot in Fig.~\ref{fig:precap} their snapshots on the $(\phi, a)$ plane at $t=-30, -20, -10$ and 0\,kyr, with $t=0$ being the moment when the nominal resonance reaches the initial position of test particles at 47.62\,AU.

As shown in Fig.~\ref{fig:precap}, the distribution of test particles in the $(\phi, a)$ plane has been disturbed by the resonance no later than $t=-30$\,kyr (about half of the resonance libration period) before the nominal resonance arrives, when the nominal resonance is about 0.15\,AU ($=5$\,AU/Myr$ \times 30$\,kyr) away.

Test particles feel the strongest perturbation from Neptune as they pass through the conjunction position (resonant angle $\phi=0^\circ$) where their semimajor axes are altered and thus deviate obviously from their original values. Although they have not yet been locked in the libration, a bunch of particles show a sign of collective motion and they gather around the leading island. The leading island will expand when the migration is removed (at $t=0$\,kyr in the simulation) and trap these objects with a high probability. Of course, if the migration continues at the same speed, much more objects will be captured by the trailing island instead because it is much bigger than the leading island (Fig.~\ref{fig:2islands}). We note that in the example shown in Fig.~\ref{fig:precap} the eccentricity (0.09) is relatively too large to be efficiently captured by the trailing island (see Fig.~\ref{fig:brakeff}), and that's why only one object (blue point in Fig.~\ref{fig:precap}) is in the trailing island finally.

\subsection{Exponential migration}

As we have shown above, the instantaneous change of migration rate may influence the outcome of resonance capture, especially alter the L/T ratio considerably. In previous simulations so far, the migration has been assumed to have constant rates (linear migration) and the rate change occurred abruptly. This is an over-simplified model, and in fact an exponential model \citepads{Malhotra1995} might be much closer to the real migration process. Meanwhile, we note that the instant migration speed in an exponential migration model is decreasing continuously, thus the slowdown effect might take place all the way. To examine this, some numerical simulations with exponential migration of Neptune are performed. 

If Neptune migrates from $a_\text{i}$ to $a_\text{f}$ in an exponential way characterized by a time scale $\tau$, the mean migration rate will be $(a_\text{f}-a_\text{i})/(2\tau)$. In our simulations presented below, Neptune migrates outward from $a_\text{i}=26$\,AU to $a_\text{f}=30$\,AU following the exponential rules with time scales $\tau= 0.1$, 0.2 and 0.5\,Myr. 

\begin{figure}[!htp]
	\centering
	\resizebox{\hsize}{!}{\includegraphics{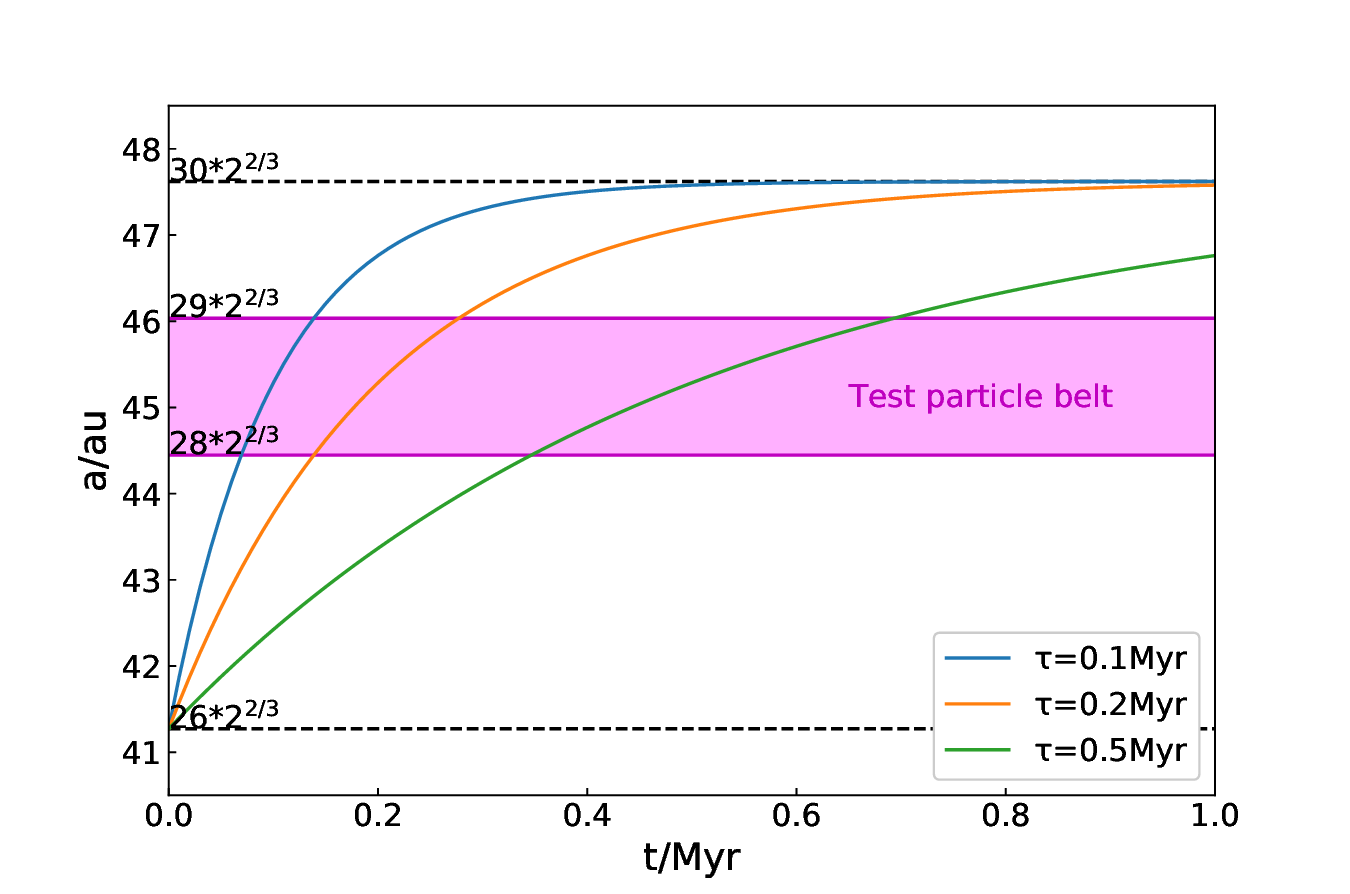}}
	\caption{The belt of test particles (shaded range) and time evolution of the nominal 1:2 resonance with Neptune.}
	\label{fig:ExpMig}
\end{figure}

As in Section~\ref{sec:basicmodel}, 10,000 test particles with eccentricities randomly in $[0,0.3]$ are located from $a_1=44.45$ to $a_2=46.03$\,AU (corresponding to the nominal 1:2 resonance when Neptune is at 28 and 29\,AU respectively), and their angular orbital elements are set randomly in $[0^\circ, 360^\circ]$. The model is sketched in Fig.~\ref{fig:ExpMig}.

\begin{figure}[!htp]
	\centering
	\resizebox{\hsize}{!}{\includegraphics{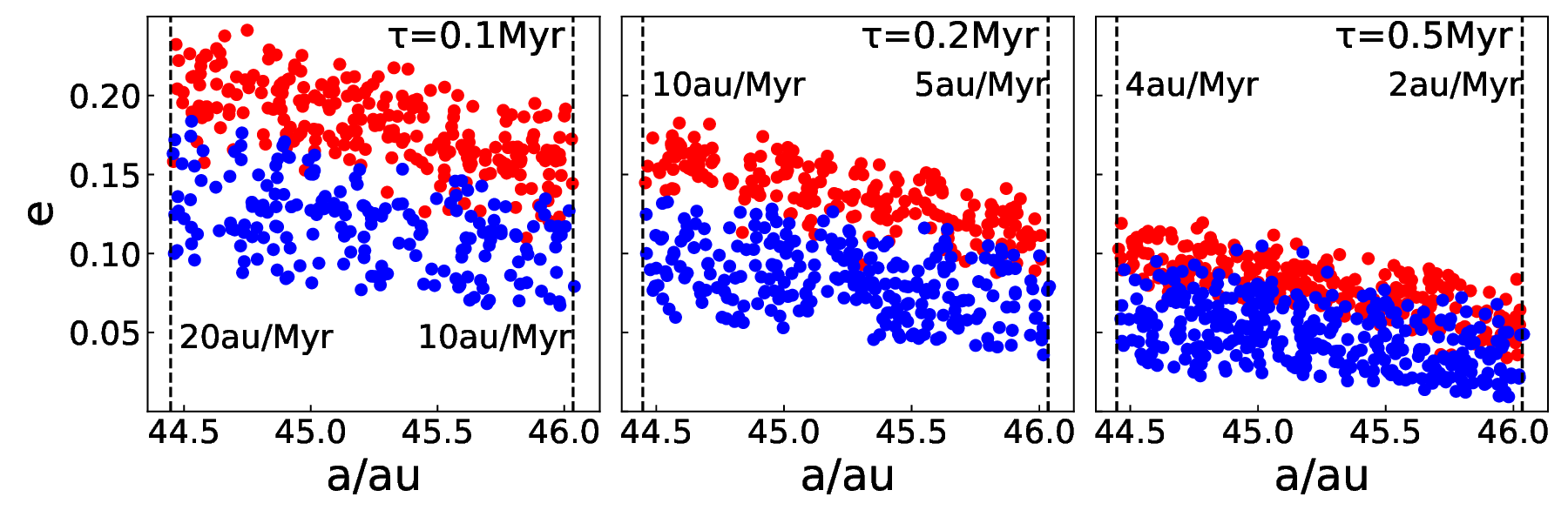}}
	\caption{Initial semimajor axes and eccentricities of objects that are captured by leading (red) and trailing (blue) islands in the exponential migration model. The time scales of migration $\tau$  are marked in each panel. The dashed lines are the inner and outer edge of the belt of test particles, and the numbers aside are the instant migration rates of Neptune when the nominal resonance arrives at both edges of the planetesimals' belt. }
	\label{fig:BrakExp}
\end{figure}

We integrate the system to $8\tau$ and check for each particle whether they are locked in any one of the resonance configurations in the end. The results are summarized in Table~\ref{tab:expcap} and Fig.~\ref{fig:BrakExp}. 

The overall efficiency of resonance capture is low, and it is a little higher when the migration is slower. This is consistent with the results in Fig.~\ref{fig:CapEff} where linear migrations are adopted. Out of the 10,000 test particles, totally 445, 462 and 583 objects have been captured in the asymmetric resonance islands for $\tau=0.1$, 0.2 and 0.5\,Myr, respectively. The overall capture efficiency is roughly comparable to the results in Fig.~\ref{fig:CapEff}. 

As for the asymmetric preference, the L/T ratio in the linear migration model (in fact with somewhat the slowdown effect, as we explained previously) increases from $\sim$0.4 to $\sim$0.8 as the migration rate increases from 1\,AU/Myr to 16\,AU/Myr (Fig.~\ref{fig:CapEff}). Because of the continuous slowdown effect in the exponential migration, the L/T ratio is remarkably higher than in the linear migration model with comparable migration speed. The L/T ratio in Table~\ref{tab:expcap} increases apparently as the migration time scale $\tau$ gets shorter (faster migration). A simple algebraic calculation reveals that the migration speed ($\dot{a}_{\text N}$) and deceleration ($\ddot{a}_{\text N}$) of Neptune are $2$\,AU/$\tau$ and $-2$\,AU/$\tau^2$ when the nominal resonance reaches the inner edge ($a_1$) of the planetesimal belt. And for the outer edge ($a_2$) these values are $1$\,AU/$\tau$ and $-1$\,AU/$\tau^2$, respectively. Thus for a migration in short time scale, the slowdown effect must be much more significant than a migration with longer time scale, and a much larger L/T ratio can be seen in the former case.

\begin{table}[htbp]
	\caption{Objects captured by the asymmetric resonance islands in the exponential migration model. The numbers of objects that are captured by the leading and trailing islands, as well as the L/T ratio are given in 2nd-4th columns. The averaged initial eccentricities of those captured objects are listed in the 5th and 6th column.   }
	\centering
	\begin{tabular}{|c|c|c|c|c|c|}
		\hline
		$\tau$ & Leading  & Trailing & L/T & Leading & Trailing \\ 
		(Myr)  & captures & captures & ratio & $\langle e_0\rangle$ & $\langle e_0\rangle$ \\ \hline
		0.1 & 253  	& 192 & 1.32  	& 0.182 & 0.118 \\ 
		0.2 & 203  	& 259 & 0.78	& 0.137 & 0.083 \\ 
		0.5 & 238  	& 345 & 0.69 	& 0.080 & 0.048 \\ 
		\hline
	\end{tabular}
	\label{tab:expcap}
\end{table}

In the work by \citetads{Chiang2002} and \citetads{Li2014}, the L/T ratio was 0.30, 0.91 and 1.1 when the migration time scale was 1, 10, and 20\,Myr respectively. Although the models adopted in their work were not identical, we see a consistent trend, that is, the L/T ratio increases as the migration rate becomes slower. This trend is opposite to our results in Table~\ref{tab:expcap} --- the faster the migration rate the larger the L/T ratio. In fact, we have seen such opposite trends in Fig.~\ref{fig:CapEff}, where the L/T ratio drops to a minimum at a migration rate of 1\,AU/Myr, starting from which both faster and slower migration will elevate the L/T ratio. 

In an exponential migration model, the migration speed changes constantly, thus all planetesimals in the belt encounter the resonance at different speeds, and they all experience the deceleration of migration. 
Thus, all phenomena observed in the migration-slowdown scenario can be identified easily here in the exponential migration (Fig.~\ref{fig:BrakExp}). For example, a faster migration captures objects with higher eccentricities, and for the same migration rate the leading island traps objects with higher eccentricities than the trailing one.

As the outward migration continues, the eccentricities of objects trapped in the resonance increase, which may drive some of them to travel from the symmetric resonance to asymmetric  island, particularly to the leading island. We note that in our simulations, the advantage of the leading island in population is enhanced further through this transferring among resonance configurations.

The timescales ($\tau$) we adopted above in Table~\ref{tab:expcap} are all quite short compared to what we expect to see in the early evolution of the Solar system. We used these short $\tau$ in our toy models to enhance the dynamical effects of the migration slowdown. If a longer, thus more comparable timescale to the ``reality'', is applied, we would expect the effects of migration slowdown will become weaker. To verify this, we carried out another simulation, where Neptune migrates outward from $a_\text{i}=23$\,AU to $a_\text{f}=30$\,AU with a timescale $\tau=3$\,Myr, which is relatively closer to the real situation. 

\begin{figure}[!htp]
	\centering
	\resizebox{\hsize}{!}{\includegraphics{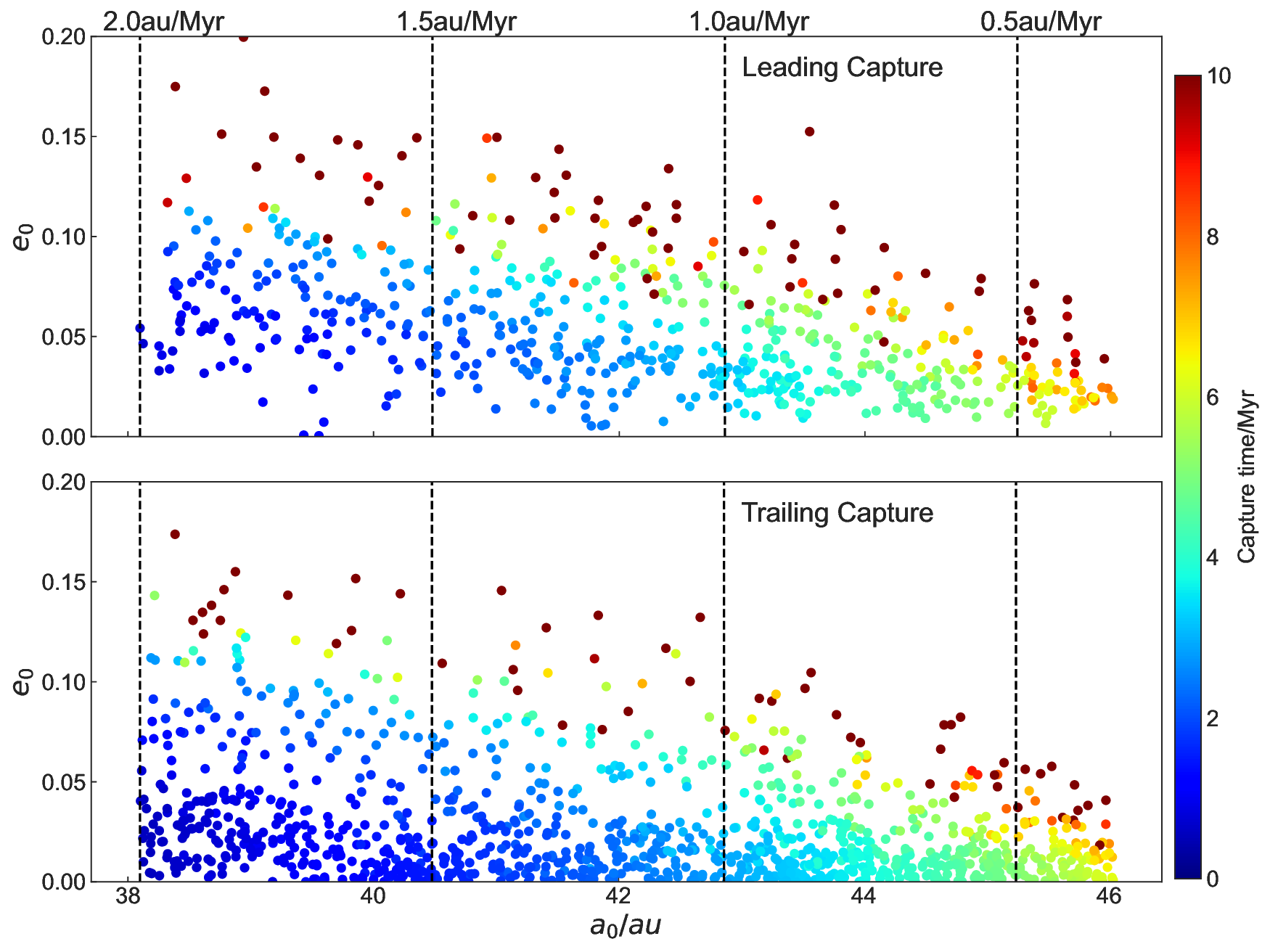}}
	\caption{Initial semimajor axes and eccentricities of objects that are captured by the leading (upper panel) and trailing (bottom panel) islands in the exponential migration model with timescale $\tau=3$\,Myr. The capture time of each object is indicated by colour.  Numbers aside dashed lines are the instant migration rates of Neptune when the nominal resonance arrives at corresponding locations.}
	\label{fig:Captime}
\end{figure}

10,000 test particles are located from $a_1=38.10$ to $a_2=46.03$\,AU (corresponding to the nominal 1:2 resonance when Neptune is at 24 and 29\,AU respectively), with their initial eccentricities randomly distributed in $[0,0.2]$, covering well the range suitable for being captured when migration is slower than 2\,AU/Myr (see Fig.~\ref{fig:CapEff}). We recorded the moment when the resonance angle crossed $180^\circ$ last time before its finally librating around either one of the asymmetric islands as the ``capture time'' for each object. The capture times of 612 leading objects and 1132 trailing objects are plotted in Fig.~\ref{fig:Captime}.

In this model, the nominal 1:2 resonance with Neptune reaches the planetesimal disk at $\sim$0.56\,Myr, and leave the disk at $\sim$6.4\,Myr. Most of the trailing objects are captured at the moment of their first contacts with the nominal resonance. On the other hand, the capture of leading objects is delayed for a long time, with a significant fraction of captures occurring after 6\,Myr, as a result of Neptune's migration slowdown. And there is a tendency that the higher the eccentricity, the later the capture time. The trailing objects also have a late-captured portion, mainly in the high eccentricity region.

 In the simulations by \citetads{Chiang2002} and \citetads{Li2014}, Neptune migrates 7\,AU, thus the average migration rates (7AU/2$\tau$)) are 3.5, 0.35, and 0.175\,AU/Myr, when $\tau=1, 10 $, and 20\,Myr, respectively. 
%Ideally, these exponential models can be compared with the linear migration model shown in Fig.~\ref{fig:CapEff}. 
In our experiment of $\tau= 3$\,Myr (Fig.~\ref{fig:Captime}), the average migration rate is $\sim$1.2\,AU/Myr ($=7\,\text{AU}/(2\times 3\text{Myr})$) and we obtained an L/T ratio of 0.54, slightly higher than the L/T ratio of 0.42 obtained in a linear migration of 1\,AU/Myr in Fig.~\ref{fig:CapEff}. However, the L/T ratio 0.30 obtained by \citetads{Chiang2002} for $\tau=1$\,Myr (average migration rate 3.5\,AU/Myr) is much lower than the L/T ratio in the linear model with comparable migration rate of 4\,AU/Myr (see Fig~\ref{fig:CapEff}). This is because the eccentricities in the former model (from 0 to 0.05) are much lower than in the latter one (from 0 to 0.3), and thus the leading island's capture of high-eccentricity objects in fast migration in our experiments did not happen in their experiment.

Therefore, in a more realistic model of longer migration timescale and of less eccentricity-excited planetesimal disk, the effects arising from the migration slowdown, such as a tuned L/T ratio (the stronger the slowdown effect, the larger the L/T ratio) and a selective capture according to eccentricity (leading island captures objects with higher eccentricity in a later stage), can be still  observed. 

In addition,  the real migration of planets caused by their encounters with planetesimals must be a significantly stochastic process  \citepads[see e.g.][]{2002MNRAS.336..520Z, Nesvorny2016}. Therefore, such acceleration/deceleration in the migration must be a common phenomenon, and the slowdown effects must leave their imprints in the distribution of objects trapped in the asymmetric resonance islands.

\section{Conclusion and discussion}

In the 1:2 mean motion resonance with Neptune, beside the symmetric configuration in which the resonant angle librates around $0^\circ$ or $180^\circ$, two asymmetric resonance islands exist beyond a certain eccentricity value (Fig.~\ref{fig:AsySol}). The one with libration centre smaller than $180^\circ$ is the leading resonance island and the other one is the trailing island. Although the dynamics of these two resonance islands are identical to each other in current planetary configuration, the populations detected in both islands show an advantage in the leading island (Table~\ref{tab:pop}). On the other hand, numerical simulations of the planetary migration and resonance capture tend to attain apparently unequal populations in these two islands, often with a preference for the trailing one (Fig.~\ref{fig:CapEff}). 

When Neptune migrates outwards, the trailing island expands while the leading island shrinks. As a result, the capture into the trailing island can be more dominant, leading to a small L/T ratio between the populations in the leading and trailing islands. The sizes of both islands are determined by the migration speed of Neptune and the initial eccentricities of planetesimals. The advantage of the trailing island is most significant when the resonance slowly sweeps through a dynamically cold belt full of planetesimals with small eccentricities (Fig.~\ref{fig:2islands} and Fig.~\ref{fig:2islsize}). 

The advantage of the trailing island may be reduced by the stickiness of the leading island, which refers to the fact that an object travels in a slower pace around the leading island (Fig.~\ref{fig:timeratio} and Fig.~\ref{fig:stickiness}). Depending on the initial eccentricity, an object may spend nearly a half to one times longer time wandering around the leading island than the trailing one, and this difference is more pronounced if the migration is faster.  This increases the possibility of an object being captured by the leading island. 

If the migration of Neptune continues at a constant speed, the competition of the aforementioned two mechanisms would still make trailing island's capture dominant (Case 2 of Table~\ref{tab:rst-brk}). But the migration has to slow down and stop at some stage. Even more, the migration is likely to be a grainy process rather than a smooth one. And we found that the acceleration and deceleration of migration may significantly alter the populations captured into the two islands, and thus the L/T ratio.  

When the migration decelerates, the leading island expands (recovering from the depressed size in  migration) and swallows planetesimals in the neighbour region, and its stickiness helps by gathering more objects around it. On the contrary, the trailing island shrinks and releasing some objects that had been captured before. This slowdown effect of migration can dramatically increase the L/T ratio. In fact, by setting an appropriate deceleration (change of migration rate), we can obtain any value of the L/T ratio (Table~\ref{tab:rst-brk}). Surely, when doing this to adjust the L/T ratio, any type of change of migration rate should be possible in physical reality. Considering the overall migration of Neptune is composed of a large number of semimajor axis shifts due to scattering events of planetesimals of a variety of sizes and masses, we still expect great possibility of migration rate changes.  

The planetesimals most affected by this migration slowdown effect are those that experience the deceleration when the capturing is ongoing (Group 2 in Fig.~\ref{fig:brakeff}). Those that have been deeply in the resonance when the slowdown occurs are influenced much less (Group 1 in Fig.~\ref{fig:brakeff}). However, if the migration is not smooth, such events of deceleration must be quite common. And in the widely adopted exponential migration model (Fig.~\ref{fig:ExpMig}), although it is smooth, the migration decelerates continuously, thus the slowdown effect works evidently (Table~\ref{tab:expcap} and Figs.~\ref{fig:BrakExp} \& \ref{fig:Captime}). 

We note that the L/T ratio can be arbitrary tuned by setting the appropriate slowdown effect. Therefore, it seems that merely the ratio of Twotinos' populations trapped in the two asymmetric resonant islands nowadays cannot serve as a good indicator of Neptune's migration history. However, if the migration always ends up with slowdown, which is a reasonable assumption, we would expect that more objects will be captured into the leading islands from the outer edge of the planetesimal belt at the final age of Neptune migration. This may leave some traces about their source regions of objects trapped in the two islands. 

Some other clues about the history of migration can be found in the eccentricities of the two populations. In a slow migration, the resonance only captures objects in nearly circular orbits, and the capture is dominated by the trailing island. Therefore, the existence of objects observed in the leading island nowadays implies that the initial orbits of planetesimals had been excited to some extent before they were captured. At a relatively higher migration rate, only those planetesimals with eccentricities in a specific range can be captured, and the leading island tends to trap objects with higher eccentricities than the trailing island. This phenomenon of leading island preferring high eccentricity gets more evident when the migration is faster. Comparing the eccentricity distributions in both asymmetric islands may also tells some important information about the origin of these objects. Solid conclusion may be made when the sample size of Twotinos is large enough someday.  

In the past, many attempts have been made to constrain the timescale of migration. Studies have explored a wide range of migration timescales, from 0.1\,Myr to 100\,Myr, yet no definitive timescale has been determined. \citetads{Murray2005} argued against fast migration, as it may lead to an overwhelming capture by the trailing island. \citetads{Nesvorny2015a} also favours an e-folding timescale greater than 10\,Myr to ensure efficient excitation of inclination. Conversely, \citetads{Volk2019} find that timescales from 5 to 50\,Myr can produce inclination distributions that match observations. \citetads{Lawler2019} suggest that the grainy slow migration model \citepads{Kaib2016}, with timescales of 30\,Myr and 100\,Myr before and after the Neptune jump, respectively, is the most consistent with observations, but confirmation of this conclusion requires additional observations.

The migration of planets in the Solar system could be a complicated procedure. For example, \citetads{Nesvorny2012} suggested that five or six giant planets ever existed in the system, and when a giant planet was ejected, Neptune’s semimajor underwent a jump. Such a major jump up to $\sim$0.5\,AU has been adopted in several subsequent models \citepads[see e.g.][]{Nesvorny2015b,Nesvorny2016,Kaib2016}. Close encounters with large planetesimals also influence Neptune, resulting in grainy migration. The variation of Neptune's semimajor axis due to close encounters with Pluto-sized objects is estimated to be $|\Delta a_N| \lesssim 0.005$\,AU, and thousands of such encountering events with massive objects may occur in a grainy migration model \citepads[e.g.][]{Nesvorny2016,Kaib2016}.

Qualitatively, a complex migration process can be a combination of a series of events simulated by simple models presented in this paper. However, as our simulations show, the outcomes of the Neptune migration and resonance capture depend sensitively on the migration model and the planetesimals' initial conditions, which makes it very difficult to reconstruct the history of planetary migration from current orbits and distribution of TNOs. 

Through a series of carefully designed numerical experiments, we found that a planetesimal around the 1:2 MMR of Neptune can ``perceive'' only the average migration rate within a time window of about 0.05 to 0.1\,Myr, which is roughly equivalent to the libration period of the 1:2 resonance angle. This time interval is much longer than the typical duration of the semimajor axis variation caused by a close encounter between Neptune and a planetesimal, which usually completes in only years. This, in fact strongly restricts our ability of retrieving the history of Neptune's migration through current distribution of Twotinos.  On the other hand, we note that a major jump of 0.5\,AU in Neptune's semimajor axis due to the ejection of a planet in such a time window would increase (or decrease) the local migration rate by approximately 5 to 10\,AU/Myr. While the average migration rate can hardly be affected by single event of close encounter with Pluto-sized planetesimal because this brings Neptune only a semimajor axis shift of $\sim$0.005\,AU. In this sense, the migration rate variation that can significantly change the L/T ratio (as shown by simulations in this paper) should be the result of planet-Neptune scattering, or arise from particular distribution of planetesimals with which Neptune exchanges angular momentum (thus migrates), rather than be merely due to the randomness of scattering events between Neptune and numerous small planetesimals.   
 
We have shown in this paper a variety of mechanisms that can regulate the L/T ratio. The mechanisms that favour the leading island's capture of planetesimals include higher eccentricity of planetesimals, migration slowdown (either abrupt or gradual), occasional inward migration, and migration rate that is much higher or lower than 1\,AU/Myr. Opposite mechanisms are in favour of trailing island's capture. All these mechanisms can cooperate to attain a more physically realistic migration model to regulate L/T ratio.

In this paper, most of the time, we adopted the planar restricted three-body problem in the numerical simulations. In fact, we have also tried a large number of simulations that include other planets and take into account the influences of orbital inclinations, both of the planets and planetesimals. We found the additional perturbations from other giant planets introduce a little more chaos in the evolution of planetesimals, but the capture efficiency of Neptune's 1:2 MMR hardly changes, and thus the conclusions about the different L/T ratios in different models are nearly the same as in the simple model presented in this paper. In addition, higher inclinations can bring changes in the resonant structure \citepads[e.g.][]{Gallardo2006, Gallardo2020, Saillenfest2016, Efimov2020}, which may break the symmetry between leading and trailing islands, or result in a sharp decrease in the capture efficiency. Of course, if the inclination is not too high, the inclination itself brings no considerable changes to the results. Considering that the original planetesimal disk is dynamically cold, a planar model still works and produces valuable information.

\begin{acknowledgements}
	Our sincere appreciations go to the anonymous referee, whose insightful comments and constructive suggestions have improved this paper greatly. This work has been supported by the science research grant from the China Manned Space Project with NO.CMS-CSST-2021-B08. We also thank the supports from the National Key R\&D Program of China (2019YFA0706601) and National Natural Science Foundation of China (NSFC, Grants No.11933001, No.12150009 \& No.12373081). 
\end{acknowledgements}

\bibliographystyle{aa-note}

\bibliography{AsyCap}

\end{document}